\documentclass[12pt]{article}
\usepackage{natbib}
\usepackage{graphicx}
\usepackage{array}
\usepackage{amsmath}
\usepackage{multirow}
\usepackage{amssymb}
\usepackage{xcolor}
\usepackage{url}

\title{Station-wise statistical joint assessment of wind speed and direction under future climates across the United States}
\author{Qiuyi Wu$^{1}$, Julie Bessac$^{2,4}$, Whitney Huang$^{3}$, Jiali Wang$^{4}$ \\
$^{1}$ {\small Department of Biostatistics and Computational Biology, University of Rochester} \\
$^{2}$ {\small Mathematics and Computer Science Division, Argonne National Laboratory, Lemont, IL, 60439, USA} \\
$^{3}$ {\small School of Mathematical and Statistical Sciences, Clemson University} \\
$^{4}$ {\small Environmental Science Division, Argonne National Laboratory}}

\date{}

\begin{document}

\maketitle

\begin{abstract}
This study develops a statistical conditional approach to evaluate climate model performance in wind speed and direction and to project their future changes under the representative concentration pathway (RCP) 8.5 scenario over inland and offshore locations across the Continental United States (CONUS). The proposed conditional approach extends the scope of existing studies by characterizing the changes of the full range of the joint wind speed and direction distribution. A von Mises mixture distribution is used to model wind directions across models and climate conditions. Directional wind speed distributions are estimated using two statistical methods: a Weibull distributional regression model and a quantile regression model, both of which enforce the \textit{circular} constraint to their resulting estimates of directional distributions. Projected uncertainties associated with different climate models and model internal variability are investigated and compared with the climate change signal to quantify the statistical significance of the future projections. In particular, this work extends the concept of internal variability to the standard deviation and high quantiles to assess the relative magnitudes to their projected changes. The evaluation results show that the studied climate model capture both historical wind speed, wind direction, and their dependencies reasonably well over both inland and offshore locations. In the future, most of the locations show no significant changes in mean wind speeds in both winter and summer, although the changes in standard deviation and 95th-quantile show some robust changes over certain locations in winter. However, in winter, high directional wind speeds are projected to decrease in Northwest, Colorado and Northern Great Plains. In summer, directional high  wind speeds over southern Great Plains slightly increase while high directional wind speeds over offshore locations do not change in future. The proposed conditional approach enables the characterization of the directional wind speed distributions, which offers additional insights for the joint assessment of speed and direction. 
\end{abstract}

\section{Introduction}\label{sec:intro}
Short-term and long-term near-surface wind variations play an important role to both human society and the environmental system ranging from wind energy \citep{Pinson09,Constantinescu11,Pinson13}; shipping industry \citep{fayle2006short,lu2013,rusu2018}; air pollution modeling \citep{zannetti2013}; wind erosion \citep{de2013wind}; building and infrastructure design \citep{mendis2007,holmes2018}; just to name a few.  
Recent California wildfires revealed that strong winds (Diablo winds in Northern California and Santa Ana winds in Southern California for instance) are a critical factor of such events \citep{westerling2004,cooley2019}. 
Therefore, assessing potential changes of both short-term variations and long-term climatology under future climate scenarios is critical and hence has received great interest in the literature \citep[e.g.,][]{pryor2009trend,mcinnes2011global,zeng2019reversal}.

Changing wind resources is of great concern for energy production and wind farm maintenance due to the varying wind magnitude and variability  \citep{pryor2010review}.
Therefore, various aspects of wind have been studied: daily wind \citep{bogardi1996daily} and wind gust \citep{cheng2014gust}. 
Studies are conducted in various parts of the world \citep{reyers2016europe,gao2018powerIndia,akinsanola2021} including North-America \citep{breslow2002powerUS,sailor2008powerUS,pryor2009trend} and the Great Lakes region \citep{pryor2009trend,li2010greatlakes}. 
Findings in the literature reveal 
regional as well as seasonal differences in wind resources, however, those results are tainted by model uncertainty and a lack of predictability of future climate that the community intends to quantify.

What makes wind different from other climate variables (e.g., temperature and precipitation) is the intrinsic vector nature that requires considering both wind speed and direction for many applications. Specifically, a wind vector can be represented in terms of the zonal (east-west) and meridional (south-north) components $(u,v)$, which is mathematically equivalently represented by the wind speed and direction $(\rho,\Phi)$. In what follows, we will use the term wind conditions to refer to both wind speed and direction. Most of the previous studies focus
on only characterizing wind speed probability distributions and their changes \citep[e.g.,][]{monahan2006probability,pryor2009trend,he2010,zeng2019reversal}. Wind direction has received less attention in the literature than wind speed due to its circular nature \citep{breckling2012} but nonetheless can be important in many applications. For instance, coastal wind direction (e.g., onshore or offshore) plays a key role in determining the magnitude of storm surge events \citep{irish2008,toro2010efficient} as well as the trajectories of storms which can potentially have severe damages when it makes landfall.
\cite{woodruff2013} showed that in coastal regions prone to tropical cyclone flooding, both wind speed and direction impact the wind-driven storm surge, which highlights the need to analyze wind speed and direction jointly. Wind direction also plays a critical role in fire spread (both prescribed and wildland fires) which can determine the affected areas \citep{abatzoglou_global_2021}. 

However, most previous studies typically consider wind speed and direction distributions separately \citep[e.g.,][]{mcinnes2011global}. A few examples that model wind speed and direction jointly are
\citep{coles1994directional}, who proposed an extreme value theory (EVT) based modeling approach for estimating directional wind speed extremes (i.e., the conditional upper tail wind speed distribution as a function of wind direction); and  \cite{pryor2012extremeEurope,de2013wind}, who considered high wind speed extremes and their associated wind directions (i.e., the conditional wind direction distribution given extreme wind speeds). 
In \citep{ailliot2015}, Markov-switching autoregressive models for bivariate time series of wind polar and Cartesian coordinates are proposed. The hidden Markov chain of these models is made non-homogeneous and depending on past wind conditions, enabling to capture the complex multimodal marginal distribution of wind speed and direction and to characterize joint temporal dynamics of speed and direction. 
\cite{bessac2016} compared various types of weather regimes (local hidden, local observed and large-scale observed) for the joint zonal and meridional components of wind in a spatiotemporal fashion. 

In this study we take a statistical model-based conditional approach to jointly model the wind speed and direction distribution. Specifically, we model wind direction using von Mises mixture distributions \citep{mardia1975modes}, a mixture model of the von Mises distribution \citep{mardia1975statistics} to preserve the circular nature. We then explore two \textit{distributional regression} \citep[see][for a review]{kneib2021} approaches, namely Weibull regression to accommodate the right skewness of wind speed distributions \citep{Brown84,monahan2006probability, solari2016} and quantile regression \citep[][see Sec.~\ref{sec:methods} for more details]{koenker1978} to estimate the conditional distribution of wind speed given the wind direction (directional wind speed distribution hereafter). Combining the estimated wind direction distribution $[\Phi]$ and the estimated directional wind speed distribution $[\rho|\Phi=\phi]$ allows one to capture the joint distribution of wind speed and direction $[\rho, \Phi]$ and hence their interactions.

In addition to the aforementioned bivariate and circular nature, wind exhibits spatio-temporal features that vary across different spatial and temporal scales e.g., spatial variation due to topography, distance to coast, temporal variation due to diurnal cycle, weather regimes, seasonality, and interannual variability. 
In particular some of these features and dependencies vary across scales, such as observed spatio-temporal correlation varying across resolutions \citep[e.g.,][]{bessac2021a}. In addition, wind speeds over large water bodies are generally stronger than over the land. As a result, offshore wind power has been growing since the 1990s enabling greater energy generation than inland farms. In 2020, the US offshore wind energy project development and operational pipeline grew to a potential generating capacity of 35,324 megawatts (MW), experiencing a 24\% increase compared to 2019 (National Renewable Energy Laboratory (NREL) Offshore Wind Database 2021).

In order to explore and characterize these features, climate model simulations are often used because of their completed space-time data structure compared to the sparsity and irregularity of observational data. Another key advantage of climate models is that they can generate future climate projections under various conditions, and therefore allows us to estimate the projected changes in wind speed and direction using our proposed method to be described in Sec.~\ref{sec:methods}.

While the outputs from general circulation models (GCMs) are generally shown to provide reasonable wind conditions at global scales, their lack of skill in simulating regional-scale phenomena has been documented extensively \citep[e.g.,][]{liang2008regional,bukovsky2011regional,gao2012projected}. Downscaling techniques are used to mitigate the low spatial resolution of GCMs through dynamical downscaling via regional climate models \citep[RCMs,][]{giorgi1999introduction} and statistical downscaling \citep{wilby1998}. RCMs account for local physical processes, such as convective and vegetation schemes, with sub-grid parameters within GCM boundary conditions for high-resolution projections. High-resolution (less than 20 km) RCMs resolve spatial and temporal dependencies much better than do GCMs \citep{wang2015model,di2012potential} and therefore can better resolve wind conditions. Meanwhile, some research have incorporated small-scale features of wind speed into the parameterizations in numerical models \citep{zeng2002, zhang2016, bessac2019, bessac2021a}. 

In this study, we use 3-hourly outputs from 12km Weather Research and Forecasting (WRF) regional climate simulations driven by three different GCMs under 10-year historical and 10-year future time-periods under a representative concentration pathway (RCP 8.5, aka the business-as-usual scenario). We evaluate climate model performance and projected future changes in wind conditions over selected inland and offshore grid cells (see Fig.~\ref{fig:map_area}) over CONUS. We consider offshore locations in shallow waters of both West and East coasts and over Lake Erie in the Great Lakes region, where the US offshore wind pipeline are installed. In addition, we analyze a 16-member RCM ensemble (with initial conditions perturbed) to quantify the effects of model's internal variability as opposed to the forced variability due to climate change.

The reminder of this paper is structure as follows: in Sec.~\ref{sec:data} we describe the RCM model outputs, benchmark reanalysis data and in-situ measurements used in this study for model evaluation; in Sec.~\ref{sec:methods}, we provide some background on the von Mises distribution, directional Weibull and quantile regression models, the bootstrap procedure we employ, and the extended versions of  the internal variability. Sec.~\ref{sec:evaluation} presents the model evaluation where reanalysis data and in-situ measurements are used to evaluate inland and offshore grid-cell locations, respectively. We assess the projected future changes in wind conditions under the RCP 8.5 scenario in Sec.~\ref{sec:projection}. The main conclusions of the work are summarized in Sec.~\ref{sec:conclusion}.

\section{Data}\label{sec:data}

\begin{figure}
    \centering
\includegraphics[width=0.7\textwidth]{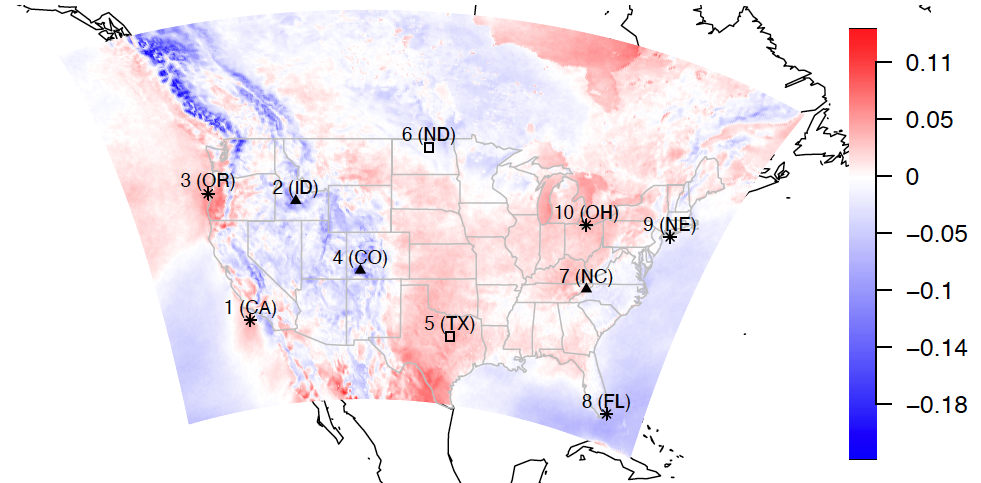}
    \caption{Studied locations: Californian coast (Location 1), Idaho inland (Location 2), Oregon coast (Location 3), Colorado mountain (Location 4), Texas Southern Great Plain (SGP) (Location 5), North Dakota Great Plain (Location 6), Southeast mountain (Location 7), Florida (Location 8), Northeast coast (Location 9), Lake Erie (Location 10). We use the ``$\ast$'' symbol to denote a offshore location, ``$\blacktriangle$'' to denote a mountain location, and ``$\square$'' to denote a plain location. The map is colored by the ratio changes in 10-year mean 95th percentile of wind speed distribution projected by WRF driven by Community Climate System Model 4 (CCSM4).}
    \label{fig:map_area}
\end{figure}

This section describes the datasets used in this study. We focus on seasonal (December-January-February (winter hereafter) and June-July-August (summer hereafter) statistics computed from the 3-hourly RCM outputs on both wind speed and direction\footnote{with meteorological conventions, meaning that the given direction is the direction from which the wind is blowing.} over ten locations with different local topological and climatological features (see Fig.~\ref{fig:map_area}). Winter and summer are chosen because of their stronger and more identifiable patterns than autumn and spring. 

\subsection{Regional climate model outputs}\label{sec:data_rcm}

We use three WRF simulations driven by Community Climate System Model 4 \citep[CCSM4,][]{gent2011community}, the Geophysical Fluid Dynamics Laboratory Earth System Model 2 \citep[GFDL-ESM2G,][]{donner2011dynamical}, and the Hadley Centre Global Environment Model version 2 \citep[HadGEM2-ES,][]{jones2011hadgem2}. These three GCMs represent a range of climate sensitivities that encompasses most of the coupled model intercomparison project phase 5 (CMIP5) GCMs when projecting future temperature changes \citep{sherwood2014spread}. For more details on these simulations, see \cite{wang2015high,zobel2018analyses,zobel2018evaluations}.

In both CCSM4- and GFDL-driven WRF runs, boundary conditions are bias-corrected using reanalysis data; and  nudging techniques are applied to WRF runs. No bias-correction or nudging are applied to the HadGEM-driven WRF runs.  
In this work, we focus on RCP 8.5 scenario for future projections, which assumes the continued heavy use of fossil fuels at a similar, or greater rate as current concentrations of CO2 and other greenhouse gases (GHGs) through the end of the century, leading to a radiative forcing of 8.5 W/$\mathrm{m}^2$ by 2100 \citep{riahi2011rcp}. For the future time period we focus on late 21st century period 2085-2094. 

A 16-member ensemble of one-year of RCM simulation using bias corrected CCSM4-driven WRF is also generated for analyzing the uncertainty due to the RCM's internal variability (IV), see methods in Sect. \ref{sec:iv}, and how significant is the climate change signal in wind speed compared with this uncertainty. These ensemble members use exactly the same model configuration such as physics parameterizations, spatial resolutions and nudging techniques, except that their initial conditions are slightly different. 
Further details about the experiment design can be found in \cite{wang2018}. 
Previous studies found that the IV was neither affected by the time period (i.e., past versus future) nor by the type of driving data  \citep[i.e., reanalysis data versus GCM output,][]{braun2012internal}. 
\cite{lucas2008investigation} found that, from a 10-member ensemble of 30-year simulations over North America, there was no long-term tendency in the IV, but there were fluctuations of the IV in time, such as in day-to-day solutions. 
Therefore, one year of ensemble simulations driven by one GCM is considered sufficient for our purpose to compare the magnitudes of future change in wind speed versus the uncertainty due to IV. See \ref{sec:iv} for details describing the calculation of IV in this study.

\subsection{Benchmark data} \label{sec:data_benchmark}

\paragraph{Reanalysis fields}
Reanalysis data are used as a verification dataset to evaluate the RCMs' wind conditions under study for the historical time period. Only a few reanalysis datasets are available at high resolution with both wind speed and wind direction. 
For the seven inland locations, we use the second phase of the  multi-institution  North  American  Land  Data  Assimilation System project, phase 2 \citep[NLDAS-2,][]{xia2012continental1,xia2012continental2}, at a spatial resolution of 12 km and hourly resolution. NLDAS-2 is an offline data assimilation system featuring uncoupled land surface models driven by observation-based atmospheric forcing. The non-precipitation land surface forcing fields for NLDAS-2 are derived from the analysis fields of the NCEP North American Regional Reanalysis \citep[NARR,][]{mesinger2006north}. NARR fields are spatially interpolated to the finer resolution of the NLDAS 1/8th-degree grid and then temporally disaggregated to the NLDAS hourly frequency. 
Since NLDAS fields are not available offshore, we use NARR fields to evaluate the RCMs' wind conditions for the offshore locations. 
NARR reanalysis fields are at a 32-km spatial resolution and 3-hourly temporal frequency.

\paragraph{In-situ measurements}
Since reanalysis data can present errors and uncertainties, ground measurements and offshore buoy measurements described below are used to consolidate the evaluation of RCMs' wind conditions for inland and offshore locations in historical climates.  
 
Observational data are extracted from the Automated Surface Observing System (ASOS) network that consists stations covers the U.S. territory. At the studied locations, measurements are recorded every minute, which are filtered over a one-hour time window in order to match the temporal resolution of NLDAS fields. 

The offshore downscaled wind speeds from the historical decade are compared with National Data Buoy Center (NDBC) buoy observations of near-surface wind velocities. The observed winds at the NBDC anemometers are adjusted to 10-m above ground height through the power-law extrapolation method \citep{Hsu1994} and collected at 3-hourly rate.

\section{Methods}\label{sec:methods}

This section describes the statistical techniques used for modeling the probability distributions of wind direction and directional wind speed. Due to the circular natural of the wind direction variable, we fit a von Mises mixture distribution \citep{fisher1995,mardia2009directional,breckling2012} to the wind directions, and we model the wind speed distributions \textit{conditioning on} wind directions by two regression models: 1) a two-step Weibull distributional regression model; 2) quantile regression \citep{koenker1978}, both of which impose a circular constraint. In the following we give a brief account for the von Mises mixture distribution, our two-step Weibull regression, and quantile regression respectively.   

\subsection{Von Mises distribution} 
The von Mises distribution (aka the circular normal distribution) has been widely used to accommodate the circular nature of wind direction. The probability density function of the von Mises distribution is given by:
\begin{equation}\label{eq: von mises}
    f(x | \mu, \kappa) = \frac{e^{\kappa \cos(x-\mu)}}{2\pi I_0(\kappa)}, 
\end{equation}
where $I_0(\kappa)$ is the modified Bessel function of order 0 \citep{hill1977algorithm}; $\mu$ is the location parameter that describes where the bulk of the ``angle'' $x$ distribution  is clustered around; $\kappa$ measures the level of concentration around the location $\mu$, the larger the $\kappa$ the more concentrate the data to $\mu$ so $\frac{1}{\kappa}$ is analogous to $\sigma^2$ in the normal distribution.

The von Mises distribution is unimodal and may lack flexibility to capture the potentially complex wind direction distribution. Therefore, we employ two-component mixture of von Mises distributions that can accommodate more complicated wind direction distributions while keeping model fitting manageable using \texttt{movMF} package developed by \citep{hornik2014} in \texttt{R}. The probability density function of the von Mises distribution mixture is given by:
\begin{equation}\label{eq: mixed von mises}
g(x|\mu, \kappa) = \pi_1 f(x|\mu_1, \kappa_1) + (1-\pi_1) f(x|\mu_2, \kappa_2),
\end{equation}
and its parameters $(\pi_1, \mu_1, \kappa_1,  \mu_2, \kappa_2 )$ are estimated via expectation–maximization (EM) algorithm \citep{dempster1977}.

\subsection{Periodic quantile regression}\label{sec:pqr}
Quantile regression (QR) extends the scope of classic regression analysis, which models the conditional \textit{mean} of a response ($\mathbb{E}(Y)$) as a function of the explanatory variables $x$'s, to modeling how a \textit{quantile} of a response $Q_{Y}(\tau) = F^{-1}_{Y}(\tau) = \inf\{y: F(y) \ge \tau\}, \tau \in [0, 1]$ changes with the explanatory variables \citep{koenker1978}. Since the quantile functions $\{Q_{Y}(\tau), \tau \in [0, 1]\}$ fully determine the distribution $F_{Y}$, one can estimate a set of conditional quantile levels  (i.e., $Q_{Y}(\tau_{k}|X=x)$, $\tau_{k} \in [0, 1], k = 1, \cdots, K$, $x \in \mathbb{R}$) to approximate the underlying conditional distribution. By doing so one can obtain a more complete picture of the full distribution of interest and how this distribution varies with the predictors \citep{mosteller1977}. More details can be found in Appendix \ref{sec: quantile regression}.

In this study we model a given quantile of wind speed varying across wind direction by representing the directional quantile curve as a periodic B-spline.  We utilize \texttt{quantreg} and \texttt{pbs} packages in \texttt{R} to implement this procedure.  
Having a collection of 
estimated conditional quantiles will provide us information on not only how a specific quantile level of wind speed change with direction but also how these quantile curves, as function of wind direction, change across quantile levels.

\subsection{Weibull distributional regression}\label{sec:WDR}
The Weibull distributional regression (WDR), an example of parametric distributional regression \citep{kneib2021} with the Weibull conditional distribution assumption, is another method we used for estimating directional wind speed quantiles. 
Under Weibull distribution assumptions, we need to estimate how the scale $\lambda$ and shape $\kappa$ parameters vary as functions of wind direction $x$ with periodicity constraints. We propose a two-stage procedure as follows: 1) we bin the data by dividing the wind direction into $N$ bins, then fit a Weibull distribution to the wind speed data within each bin via maximum likelihood method to obtain the estimates $\{\hat{\lambda}_{j}, \hat{\kappa}_{j}\}_{j=1}^{N}$ and their standard error $\{se(\hat{\lambda}_{j}), se(\hat{\kappa}_{j})\}_{j=1}^{N}$;    
2) Estimate $\lambda(x)$ and $\kappa(x), x \in [-\pi, \pi]$
via a harmonic regression (i.e., $\lambda(x) =\sum_{k=1}^{K}\left\{\alpha_{\lambda, k}\cos(x \times k)+\beta_{\lambda, k}\sin(x \times k)\right\}$, and $\kappa(x) =\sum_{k=1}^{K}\left\{\alpha_{\kappa, k}\cos(x \times k)+\beta_{\kappa, k}\sin(x \times k)\right\}$). Specifically, with a chosen $K$, we fit by weighted least squares to $\{\hat{\lambda}_{j}, \tilde{x}_{j}\}_{j=1}^{N}$ and $\{\hat{\kappa}_{j}, \tilde{x}_{j}\}_{j=1}^{N}$ to get $\{\hat{\alpha}_{\lambda, k}, \hat{\alpha}_{\kappa,k}, \hat{\beta}_{\lambda,k}, \hat{\beta}_{\lambda,k}\}_{k=1}^{K}$  where the weights are the reciprocal of squared standard errors and $\tilde{x}_{j}$ is the average wind direction within the $j_{th}$ bin. The use of harmonic regression here ensures both $\lambda(x)$ and $\kappa(x)$ to be circular functions.

\subsection{Quantify estimation uncertainty via bootstrap}\label{sec:boot}

\begin{figure}
\centering
\includegraphics[scale= 0.28]{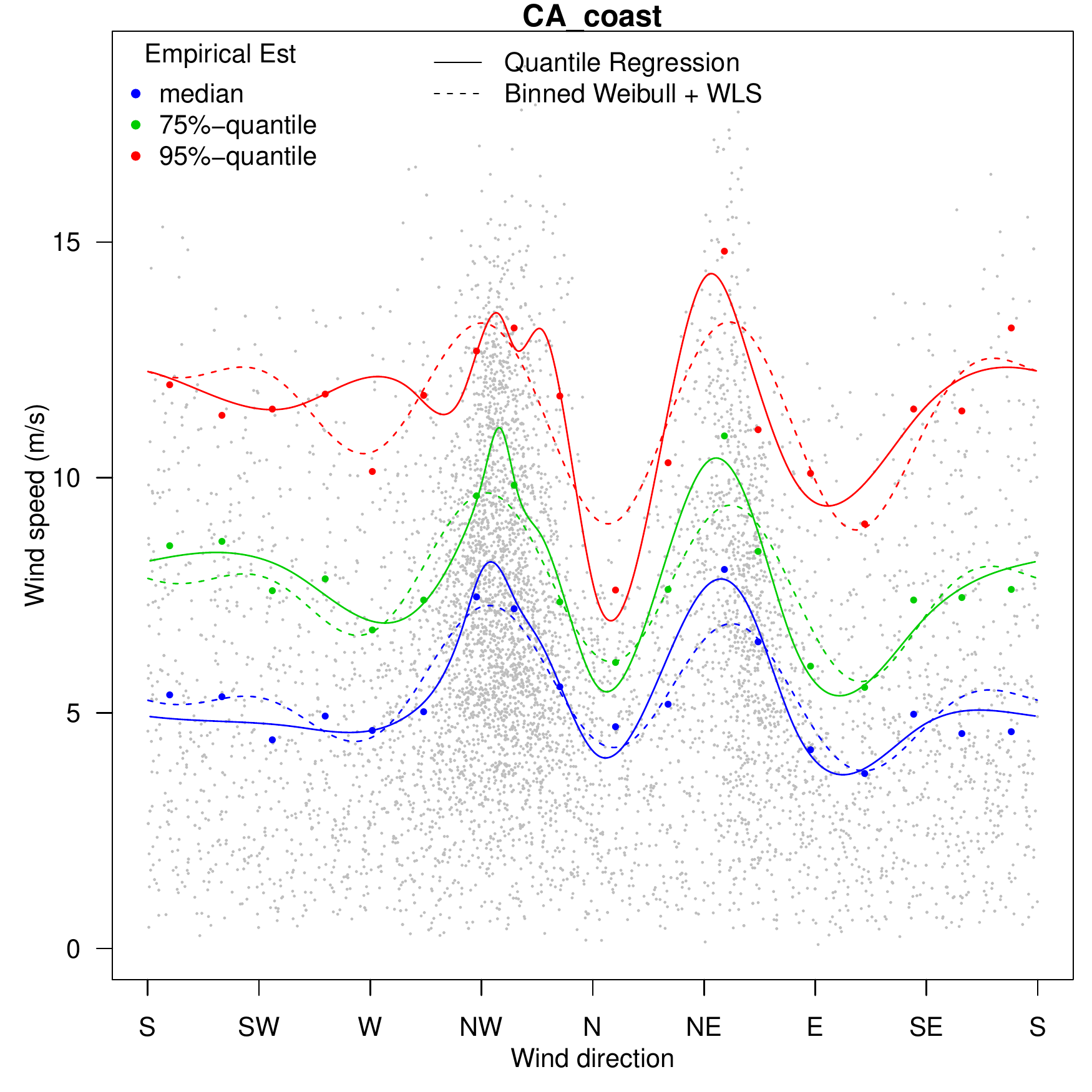}
  \includegraphics[scale= 0.3]{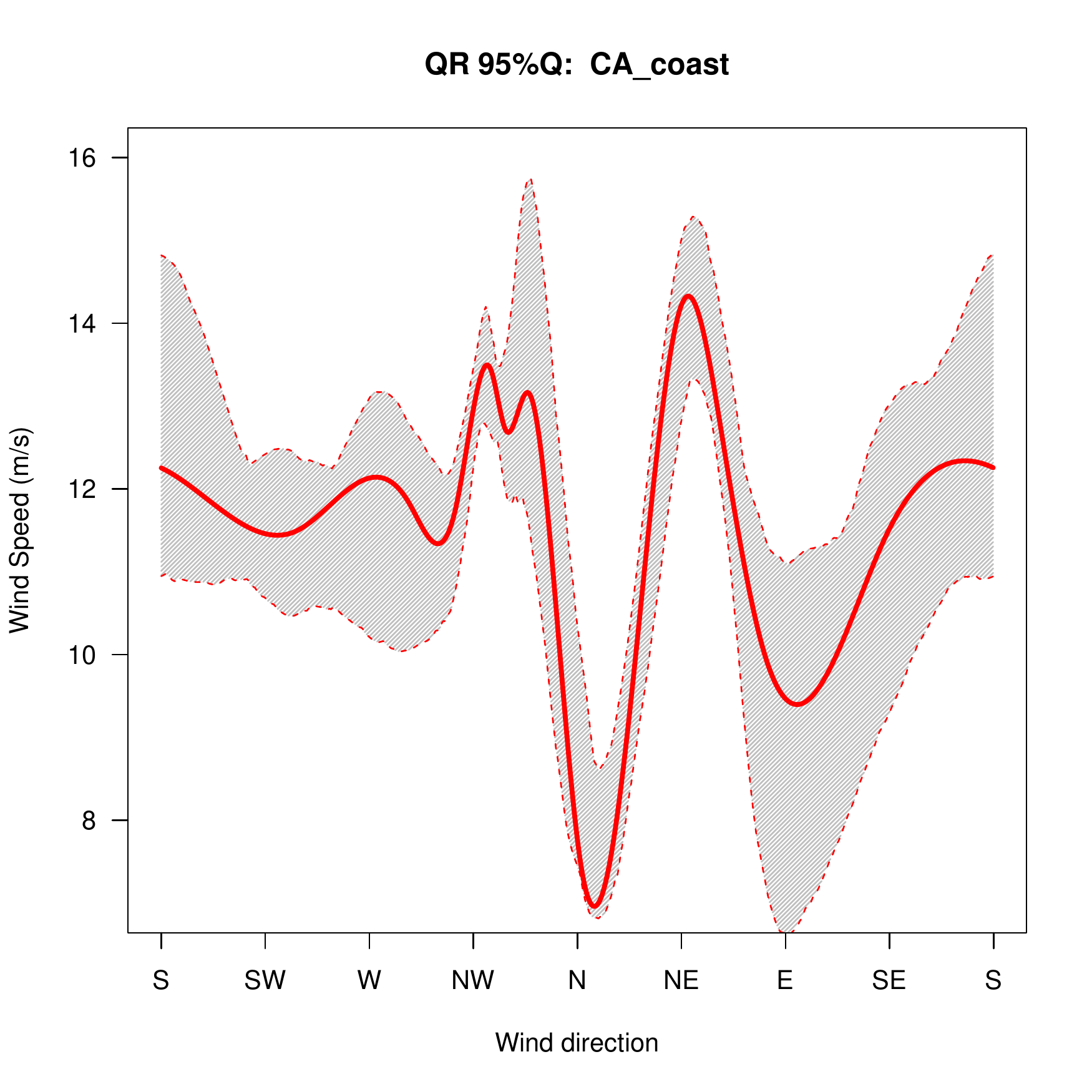}
    \includegraphics[scale= 0.3]{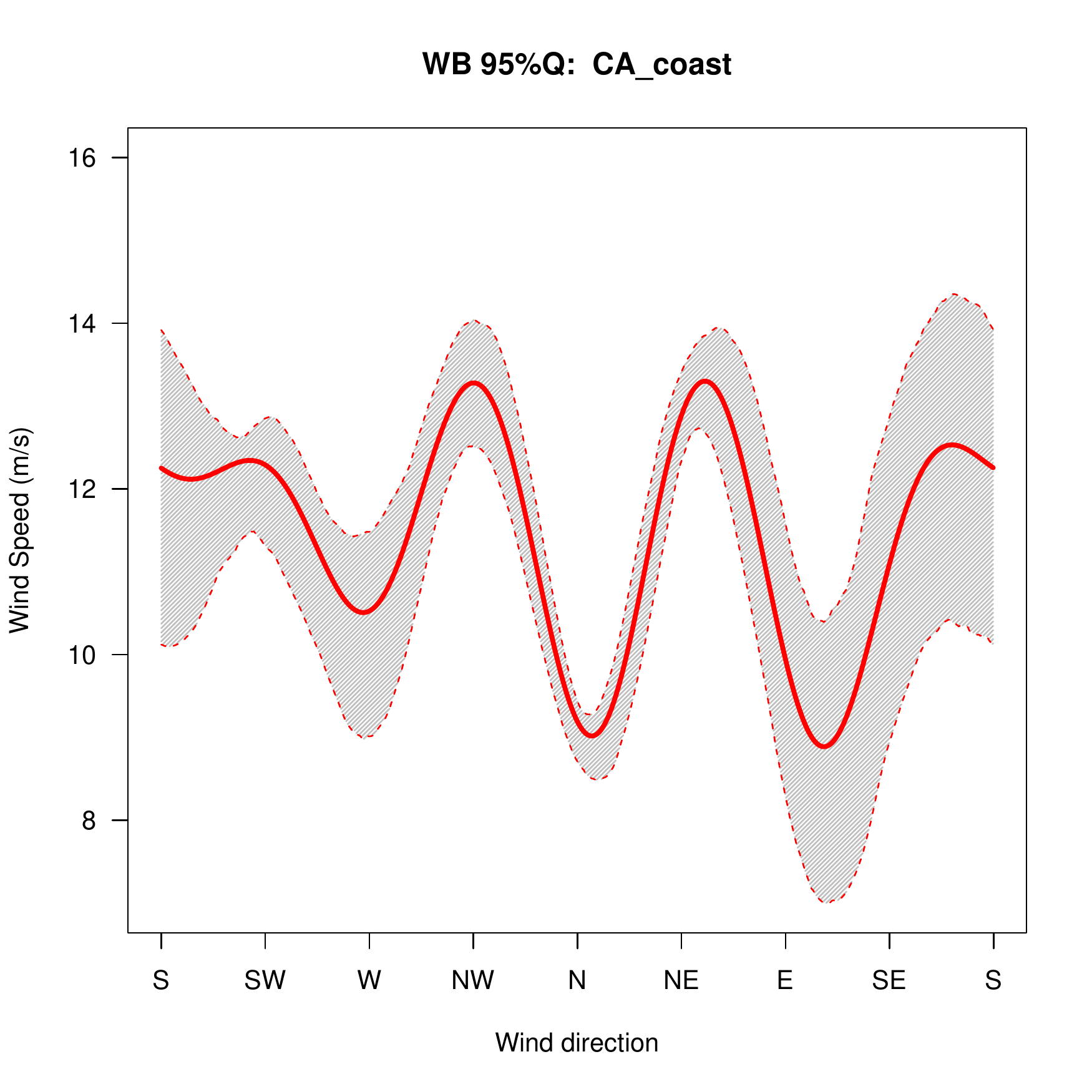}
 \caption{WRF-CCSM winter California coast scenario. \textbf{Left}: Fitted quantile curves (50th, 75th, and 95th-quantiles) of the quantile and Weibull regression methods at the CA grid point, along with their binwise empirical estimates. \textbf{Middle}: 95\% bootstrapped confidence intervals of 95th-quantile (gray shaded area) and its ``point estimate'' (red curve) using periodic quantile regression. \textbf{Right}: Same as the middle panel but using Weibull regression.}\label{fig:uq}
\end{figure}

For both QR and WDR, we use bootstrapping  \citep{efron1994} to quantify the uncertainty associated with the parameter estimation. 
Specifically, we draw 500 block-bootstrap samples, where the block size is taken to be one season to preserve the temporal dependence within a season. The $\alpha/2$ upper and lower percentiles of the bootstrap distribution are used to form a $100\times(1-\alpha)\%$ confidence interval to quantify estimation uncertainty (see Figure \ref{fig:uq} for an example). In Sect. \ref{sec:projection} and Supplementary Material Sect. 3, we provide $100\times(1-\alpha)\%$ confidence intervals for other selected locations.

\subsection{Internal variability and projected climate change signal} \label{sec:iv}

In this section, we propose new statistics based on the commonly used internal variability (IV). 
IV  arises from intrinsic variations of the nonlinear physical and dynamical processes that are described by climate models \citep{hawkins2009,wang2018}. Due to the restrictions on the large-scale atmospheric flow imposed by the lateral boundary conditions, the level of IV generated by RCMs is smaller than those generated by GCMs (at least at the large scale). However, it is important to evaluate the IV of an RCM because this variability may modulate or even mask physically forced signals in the model \citep{braun2012internal,deser2012communication}.  
In order to assess the strength of climate change signals relatively to the internal variability of the system, we compare the intensity of the climate change signal to the IV of the WRF model.  
The IV is typically measured by the time-wise spread across members of an ensemble averaged over given time-windows. 
Members of an ensemble are commonly started with perturbed initial conditions.

The IV usually represents the average over a time window $\mathcal{T}$, where $\mathcal{T}$ represents days, months or seasons, of the time-wise ensemble spread: 
\begin{equation}
 IV_{mean}(i,j,\mathcal{T}) = \left(\mathbb{E}_{t \in \mathcal{T}}\left[\frac{1}{N}\sum_{n=1}^N\left[Y_n(i,j,t)-\frac{1}{N}\sum_{n=1}^N Y_n(i,j,t)    \right]^2\right]\right)^{\frac{1}{2}}
\end{equation}

where $Y_n(i,j,t)$ refers to a variable $Y$ on grid point $(i,j)$ at time $t$ and the member $n$ in the $N$-ensemble. 
In this study, we consider $N = 10$ members described in Sect. \ref{sec:data_rcm} and compute the IV over the summer and winter time-windows for 6-hourly data. 

The commonly used IV provides information on the ensemble variability with respect to the mean of the quantity of interest, however it does not provide information regarding the internal variability of other summary statistics, such as its variance or quantiles. 
This study extends the concept of IV to compute the ensemble spread of the standard deviation and 95th-quantile of the wind speed. 
For the IV of the standard deviations, we calculate the spread across ensemble members of the wind speed standard deviation: 
\begin{equation}
 IV_{sd}(i,j,\mathcal{T}) = \left(\frac{1}{N}\sum_{n=1}^N\left[\sigma_n(i,j,\mathcal{T})-\frac{1}{N}\sum_{n=1}^N \sigma_n(i,j,\mathcal{T})  \right]^2 \right)^{\frac{1}{2}}  
\end{equation}
where $\sigma_n(i,j,\mathcal{T})$ is the standard deviation of the ensemble member $n$ at location $(i,j)$ for the season $\mathcal{T}$, 
\begin{equation}
 \nonumber    \sigma_n(i,j,\mathcal{T}) =  \left(\frac{1}{\mathcal{T}}\sum_{t=1}^\mathcal{T}\left[Y_n(i,j,t)-\bar{Y}(i,j,.)  \right]^2\right)^{\frac{1}{2}}. 
\end{equation}
where $\bar{Y}(i,j,.) = \frac{1}{N}\sum_{n=1}^N Y_n(i,j,\cdot).$
Similarly, for the 95th-quantile IV, we calculate the 95th-quantile value of each ensemble member $n$ over a season $\mathcal{T}$, and then calculate the ensemble spread of the member 95th-quantiles, 
\begin{equation}
 IV_{q95}(i,j,\mathcal{T}) =\left(\frac{1}{N}\sum_{n=1}^N\left[Q_n(i,j,\mathcal{T})-\frac{1}{N}\sum_{n=1}^N Q_n(i,j,\mathcal{T})  \right]^2\right)^{\frac{1}{2}}
\end{equation}
where $Q_n(i,j,\mathcal{T})$ is the  95th-quantile of the ensembles over a season $\mathcal{T}$.

When  projected climate changes (PCC), which are derived as the difference between a historical and projected statistics, are two-fold larger than the internal variability, changes can be considered robust with respect to the IV \citep{wang2018}.  
These  computations are  performed on WRF-CCSM ensemble described in Sect. \ref{sec:data_rcm}.

\section{Evaluation of wind conditions}\label{sec:evaluation}
In this section, we evaluate the historical RCM outputs against reanalysis data and in-situ measurements as described in Sect. \ref{sec:data} via different statistical methods described in Sect. \ref{sec:methods}.

\subsection{Comparison of reanalysis benchmark data and in-situ measurements}

\begin{figure}
  \centering
    \includegraphics[scale=0.5]{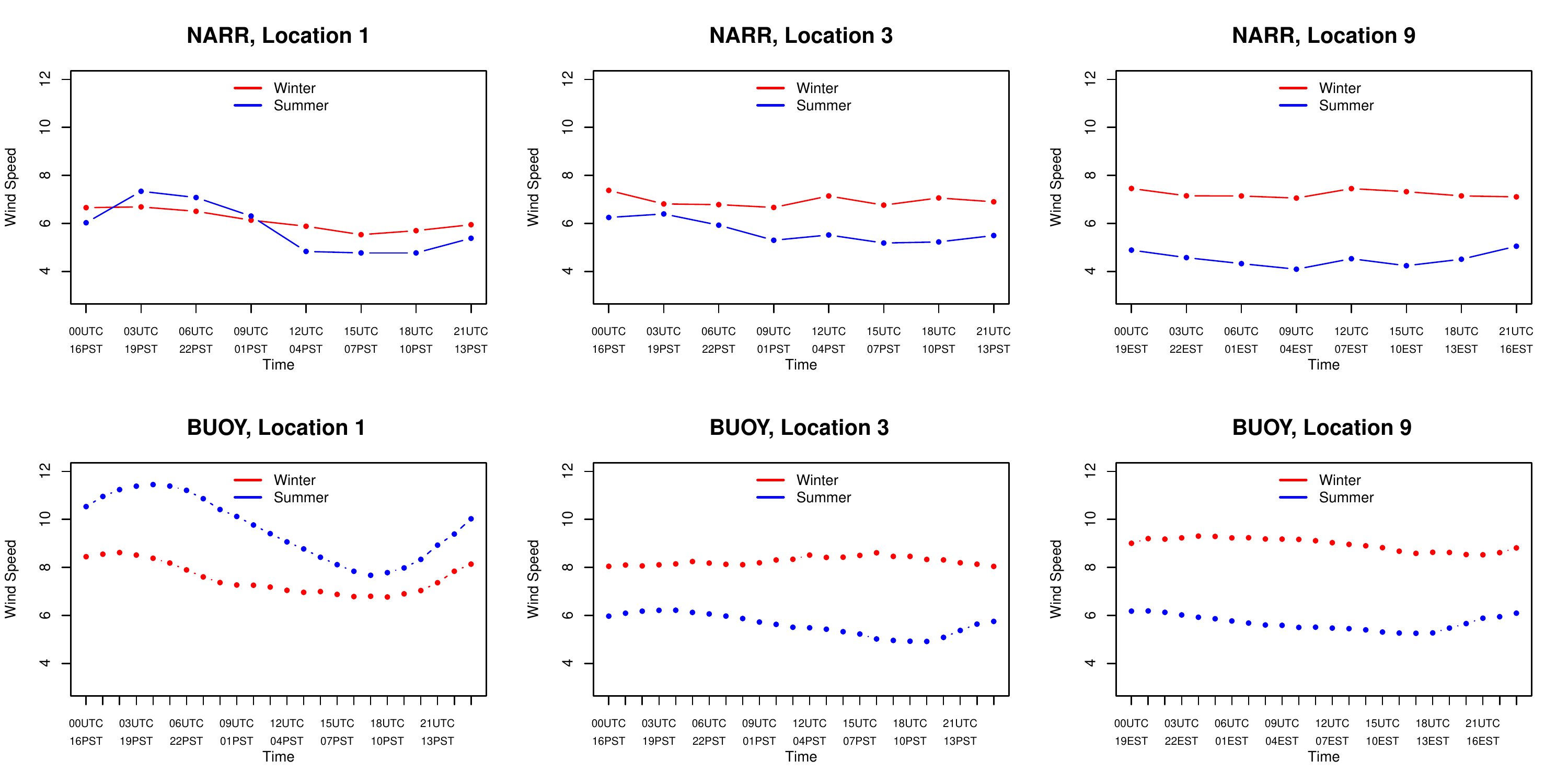}
    \caption{Diurnal wind speed pattern of NARR (\textbf{top}) and buoy (\textbf{bottom}) wind data for the three offshore locations in CA coast, NW coast, NE coast from left to right, displayed in  winter (blue) and summer (red).}
    \label{fig:diurnal1}
\end{figure}

First, we evaluate the reanalysis benchmark data NARR using in-situ buoy observation over coastal locations, and the reanalysis benchmark data NLDAS using near-surface observation ASOS over inland locations. 
Figure \ref{fig:diurnal1} shows the diurnal patterns of wind speed for three offshore locations (CA coast, NW coast and NE coast) of NARR reanalysis and buoy measurements. 
NARR captures the seasonal differences between summer and winter wind speed shown in buoy data e.g., winter has stronger wind speed than does summer in offshore locations in northwest and northeast. The diurnal pattern of wind speeds over CA coast is also captured by NARR. 
However, for both summer and winter wind speed, NARR tends to underestimate the wind speeds compared with buoy measurements. 

\begin{figure}
  \centering
    \includegraphics[scale=0.4]{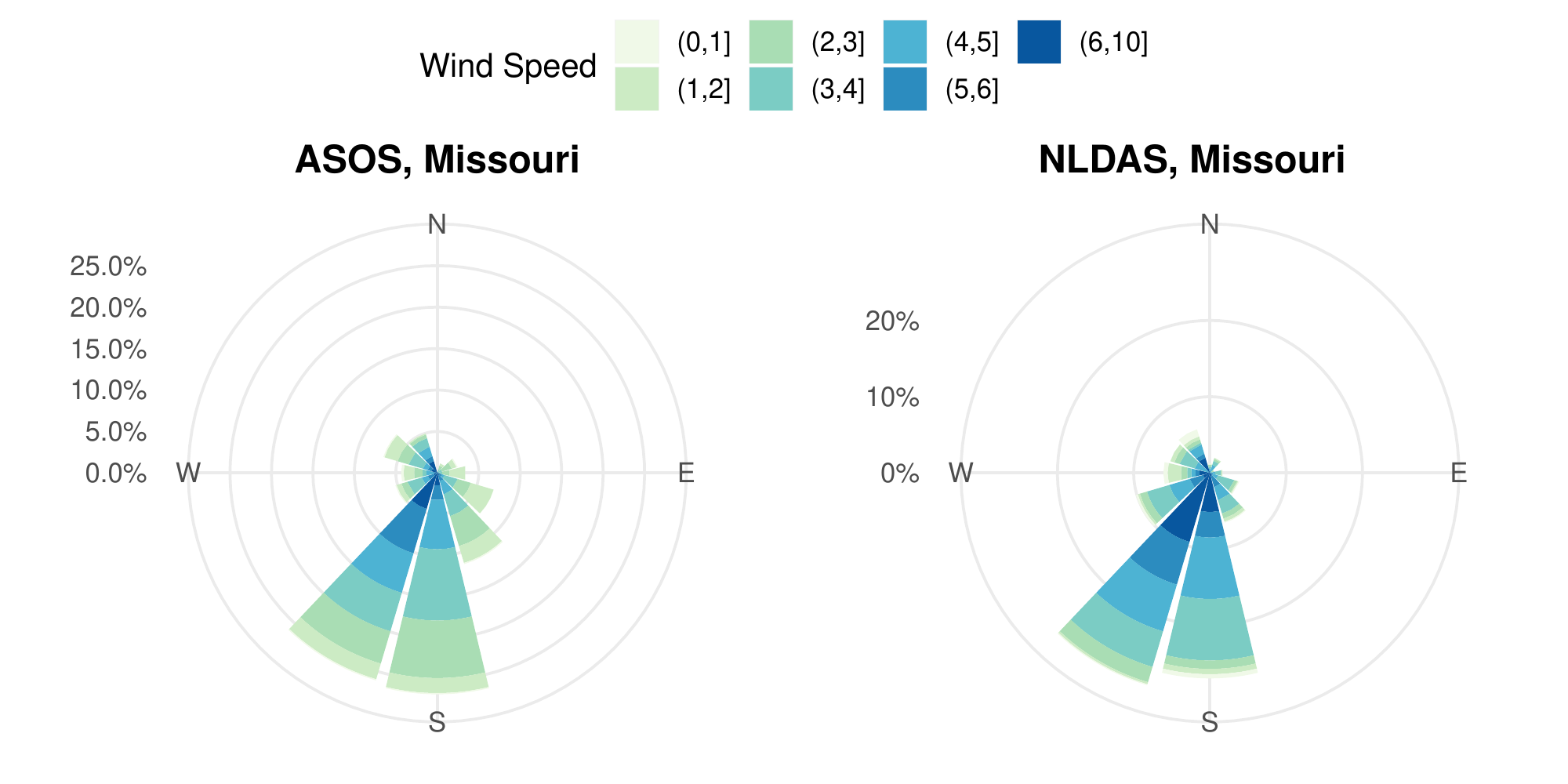}
    \includegraphics[scale=0.4]{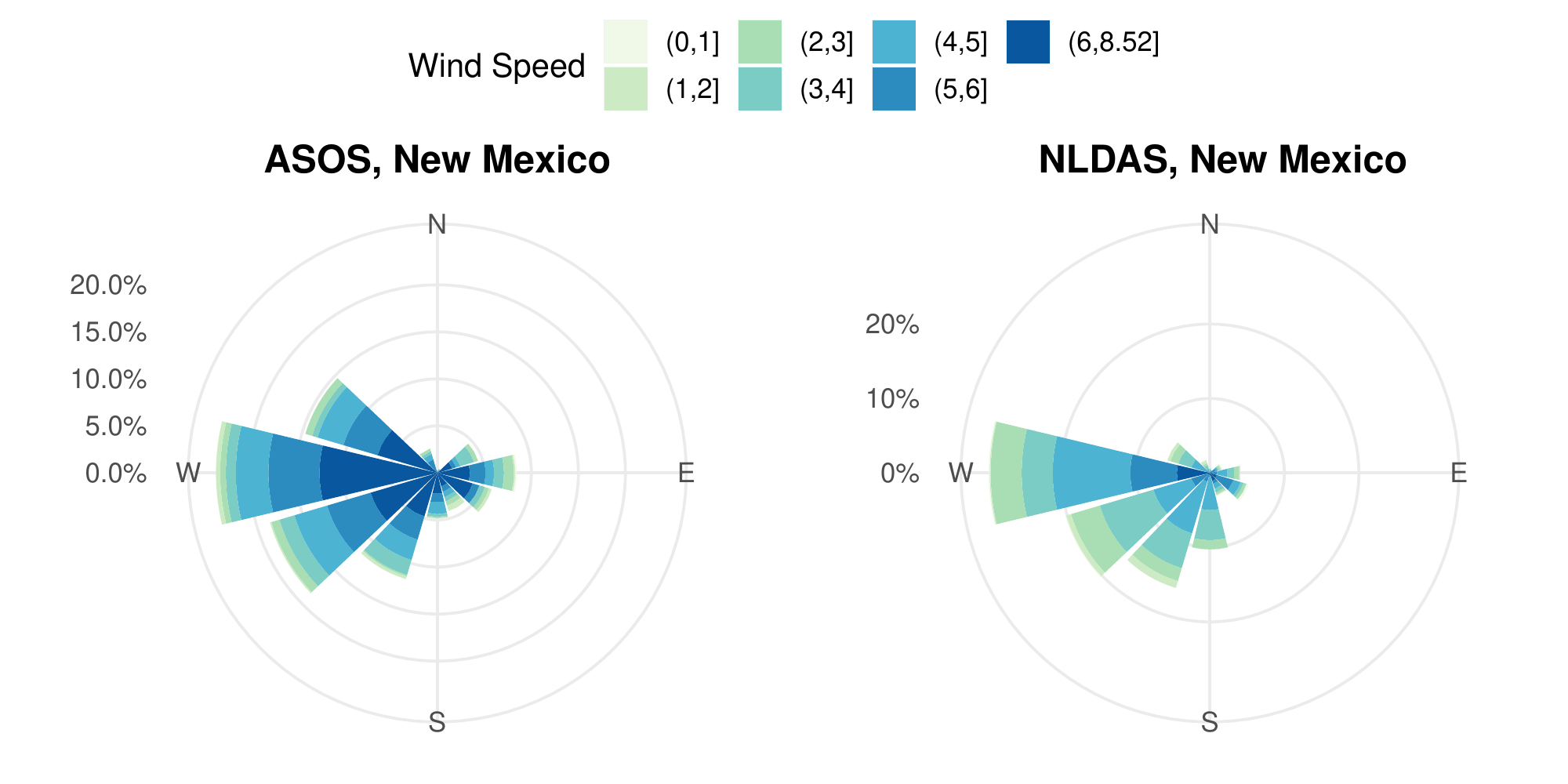}
    \caption{Windrose of ASOS and NLDAS data for two inland locations: Missouri  (\textbf{left}) and New Mexico (\textbf{right})} 
    \label{fig:ASOS vs NLDAS}
\end{figure}

Figure \ref{fig:ASOS vs NLDAS} shows windroses from ASOS and NLDAS data, using two arbitrary locations over the US in 2018 June where strong low-level jets occurred frequently over nights. Overall, we find that NLDAS reanalysis data captures both wind speed and direction fairly well. We have also evaluated other locations across the CONUS, and find that NLDAS is a reasonable benchmark data for the selected inland locations. 

\subsection{Wind direction evaluation}
\begin{figure}
 \includegraphics[scale= 0.43]{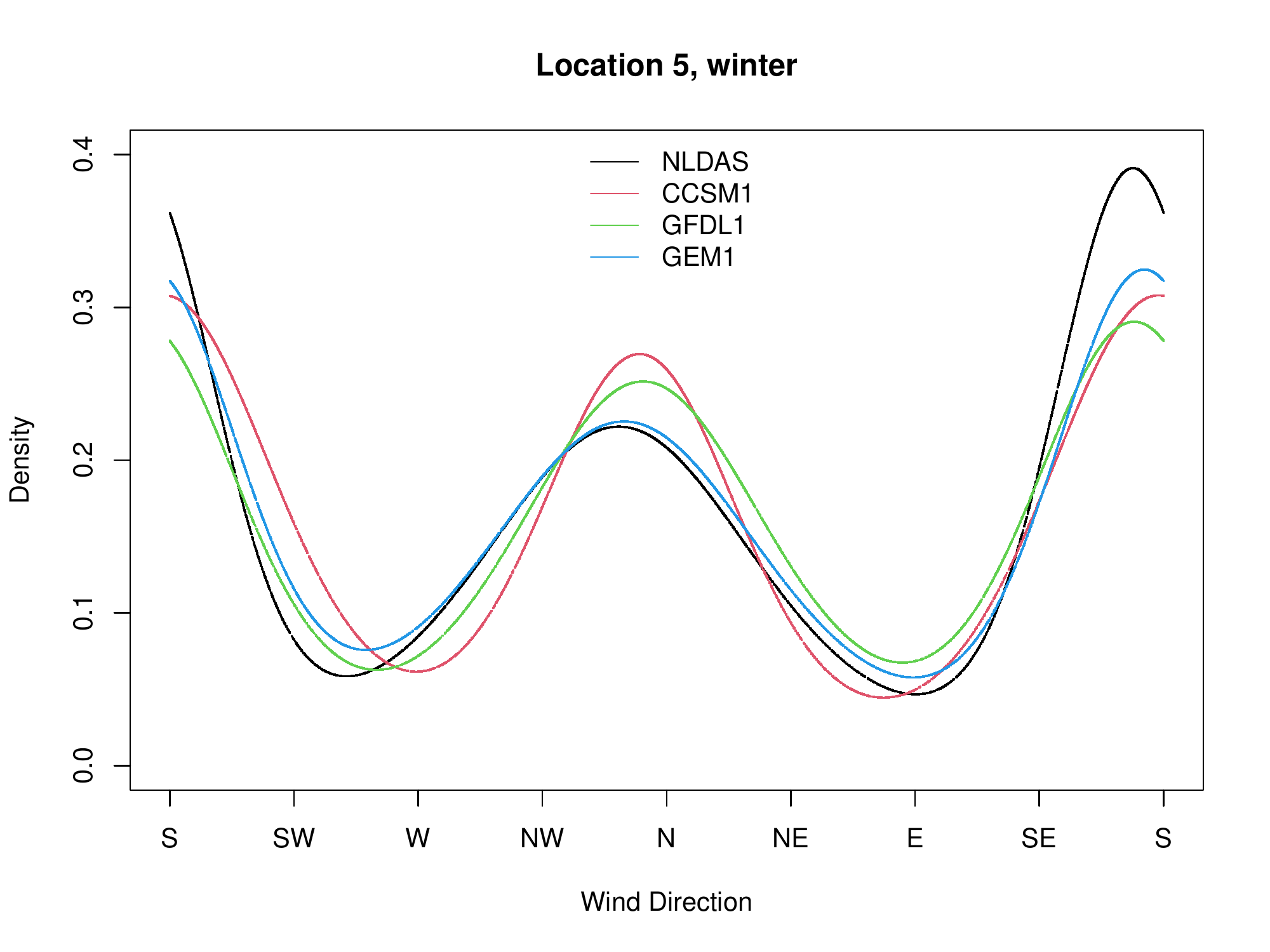}
 \includegraphics[scale= 0.43]{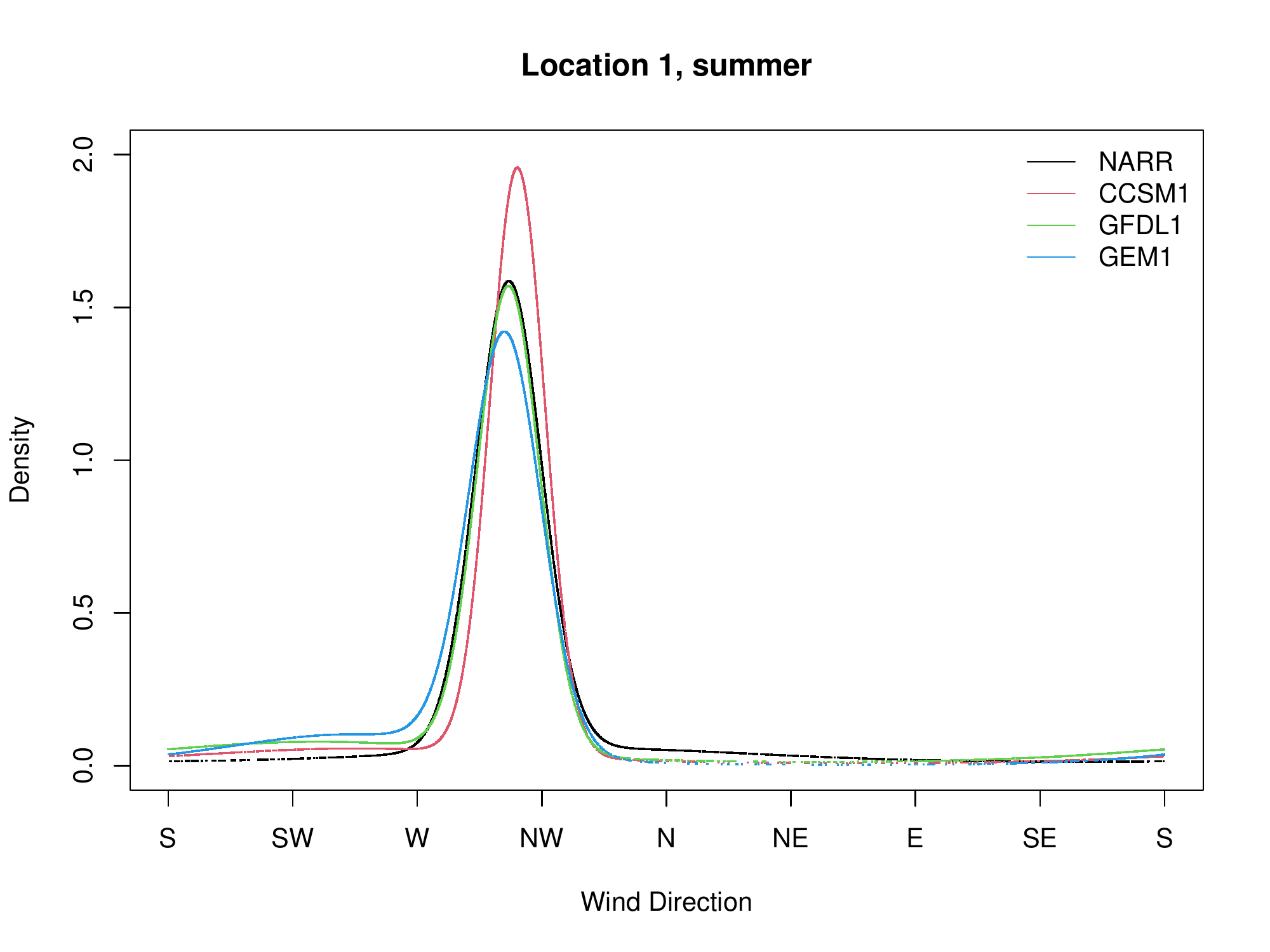}
 \caption{Fitted von Mises mixture densities for historical data. \textbf{left}: inland Texas-SGP winter, \textbf{right}: offshore CA summer. The black, red, green, blue curves represent benchmark (NARR for offshore and NLDAS for inland), WRF-CCSM, WRF-GFDL, WRF-HadGEM models.}\label{fig:vm2}
 \end{figure}

 Figure \ref{fig:vm2} shows von Mises density estimates for inland Texas-SGP location in winter (left) and offshore location CA coast in summer (right). 
 The estimated von Mises densities for the remaining locations can be found in the supplementary materials. In general, all three WRF simulations capture the variability of wind direction across all the locations reasonably well. This finding is also consistent to Figures \ref{fig:qr1.win} and \ref{fig:qr1.sum}. 
 The inland location Texas-SGP  in winter shows two dominant wind directions (southerly and northerly), hence the mixture of von Mises distributions appears well suited. The three WRF models are all consistent among themselves, as well as similar to the benchmark NLDAS data results. In summer, the offshore location CA coast shows a strong concentrated north-western wind, and all the WRF models are similar to the benchmark results. 
 Wind direction in both summer and winter for WRF models and benchmark data are similar except some locations. In addition, WRF-CCSM and WRF-HadGEM show more consistency than WRF-GFDL model in some locations in winter. From Figure \ref{fig:vm2}, we can see inland locations such as Texas-SGP has more dispersed winter direction than offshore locations such as CA. Both scenarios have very consistent patterns in all WRF models and NLDAS benchmark data.
 
Overall, compared with winter, summer winds are more concentrated around their dominant directions. CA coast, NW inland, CO mountains, NGP, SE mountains, Lake Erie have winds mainly from the same direction in both winter and summer. In particular, CA coast, NW coast, Texas-SGP have very concentrated wind while the others are quite dispersed.

\subsection{Directional wind speed distribution}
\begin{figure}
 \centering
  \includegraphics[scale= 0.3]{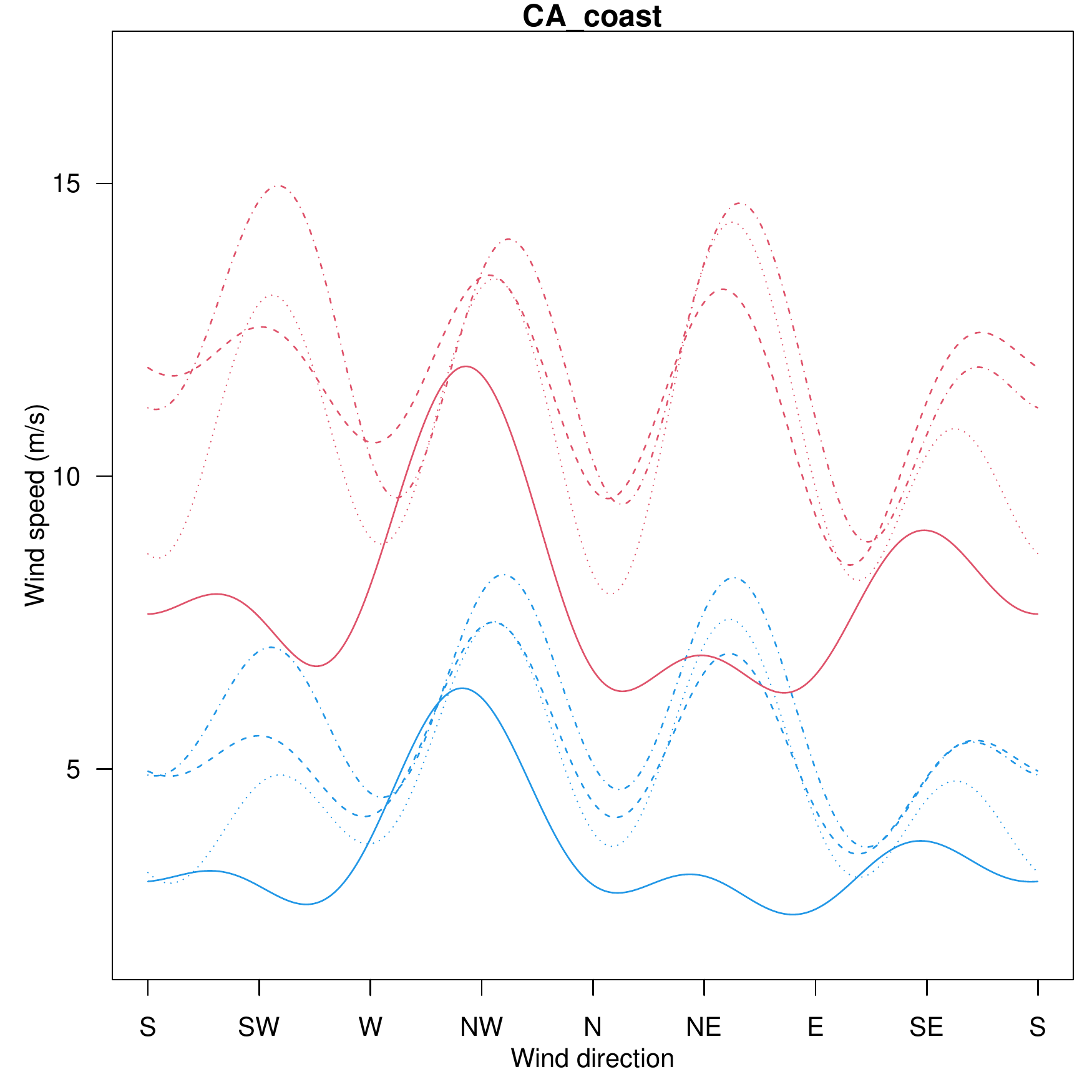}
   \includegraphics[scale= 0.3]{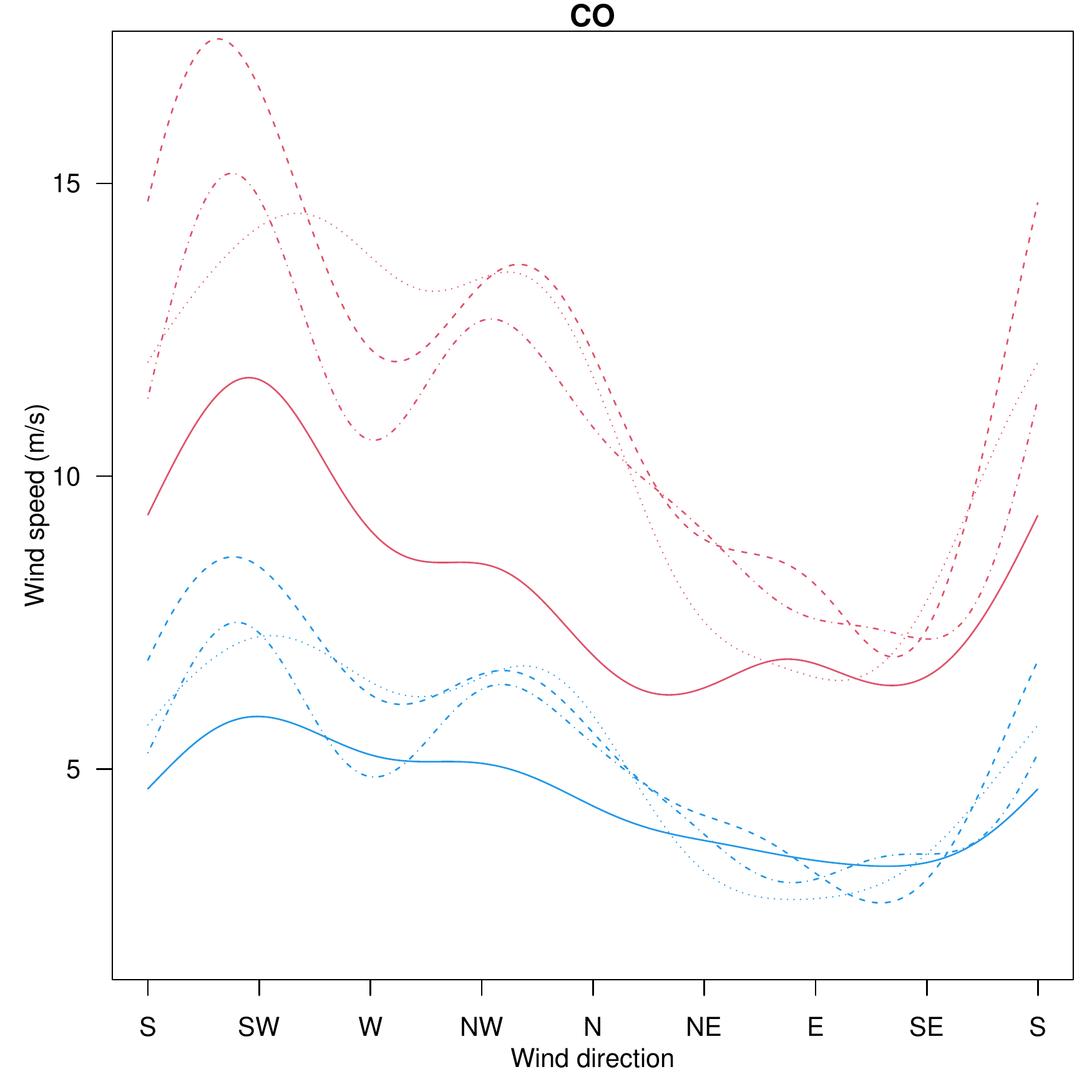}
  \includegraphics[scale= 0.3]{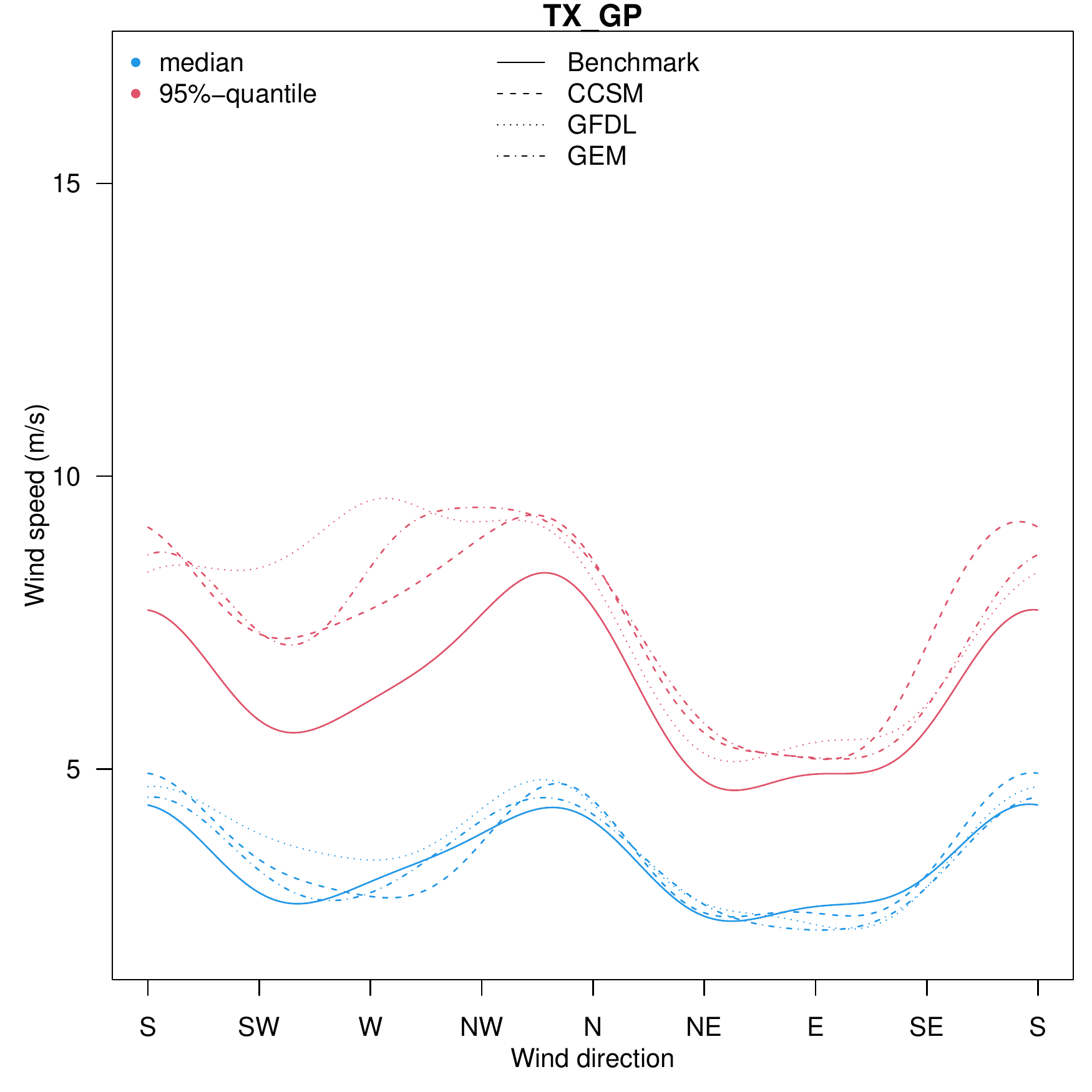}
 \caption{50th (blue) and 95th (red) quantiles of wind speed as a function of wind direction from the binned Weibull WLS model for historical period in winter for three locations, \textbf{from  left to right}: offshore California, Colorado and Texas-SGP. The different RCMs are shown with different types of dotted and dashed line, benchmark data are in solid lines (NARR for offshore and NLDAS for inland).}  
 \label{fig:qr1.win}
 \end{figure}

\begin{figure}
  \centering
 \includegraphics[scale= 0.3]{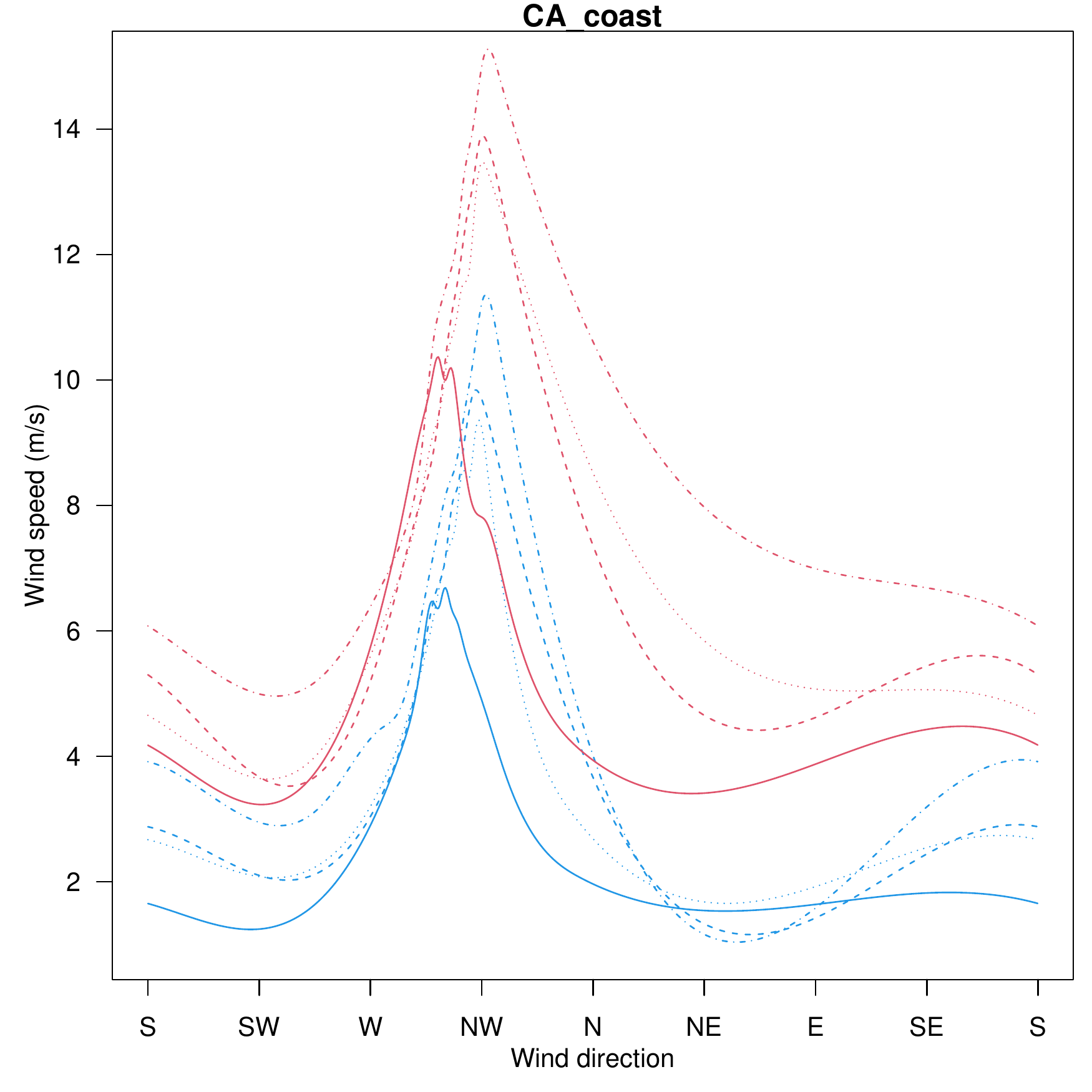}
 \includegraphics[scale= 0.3]{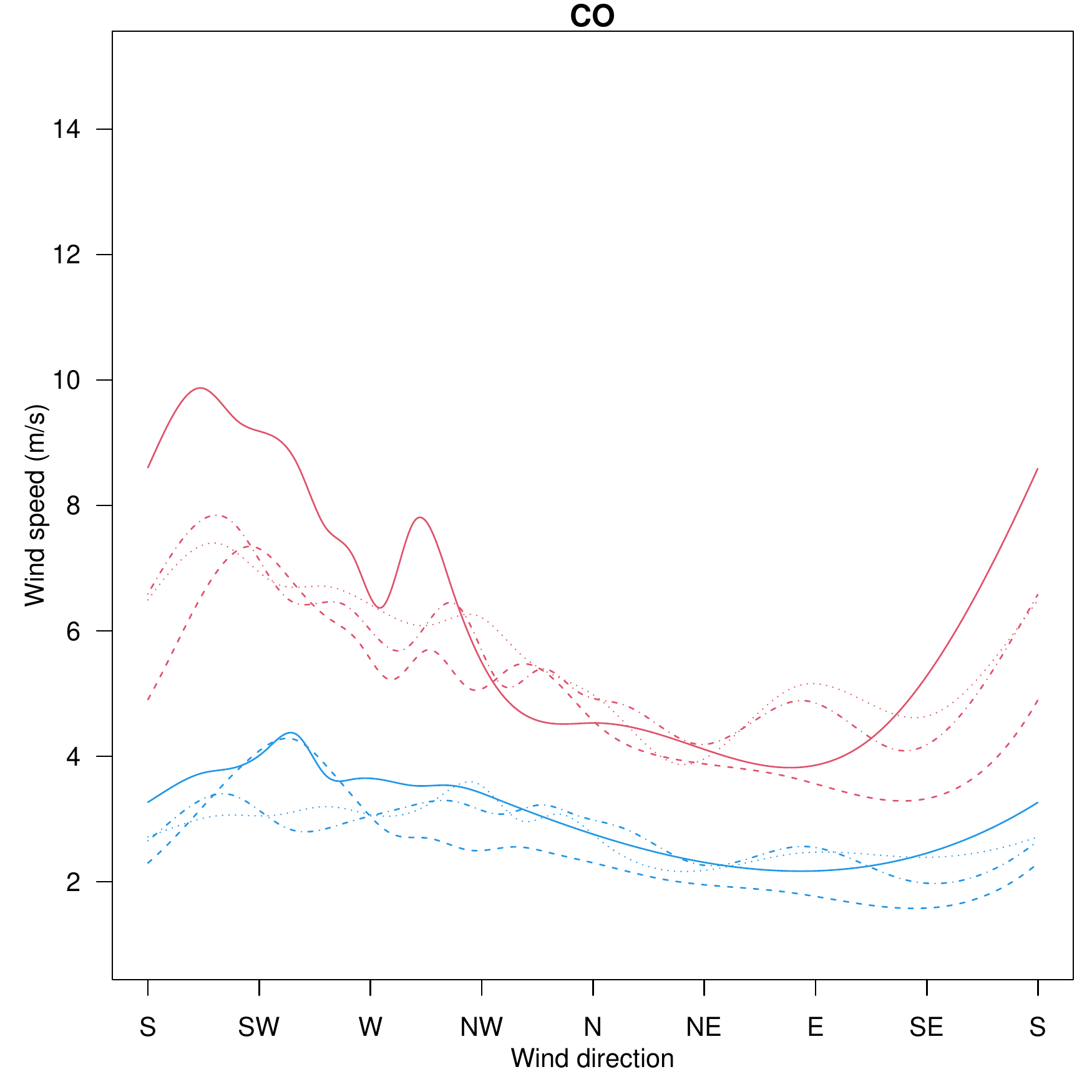} 
 \includegraphics[scale= 0.3]{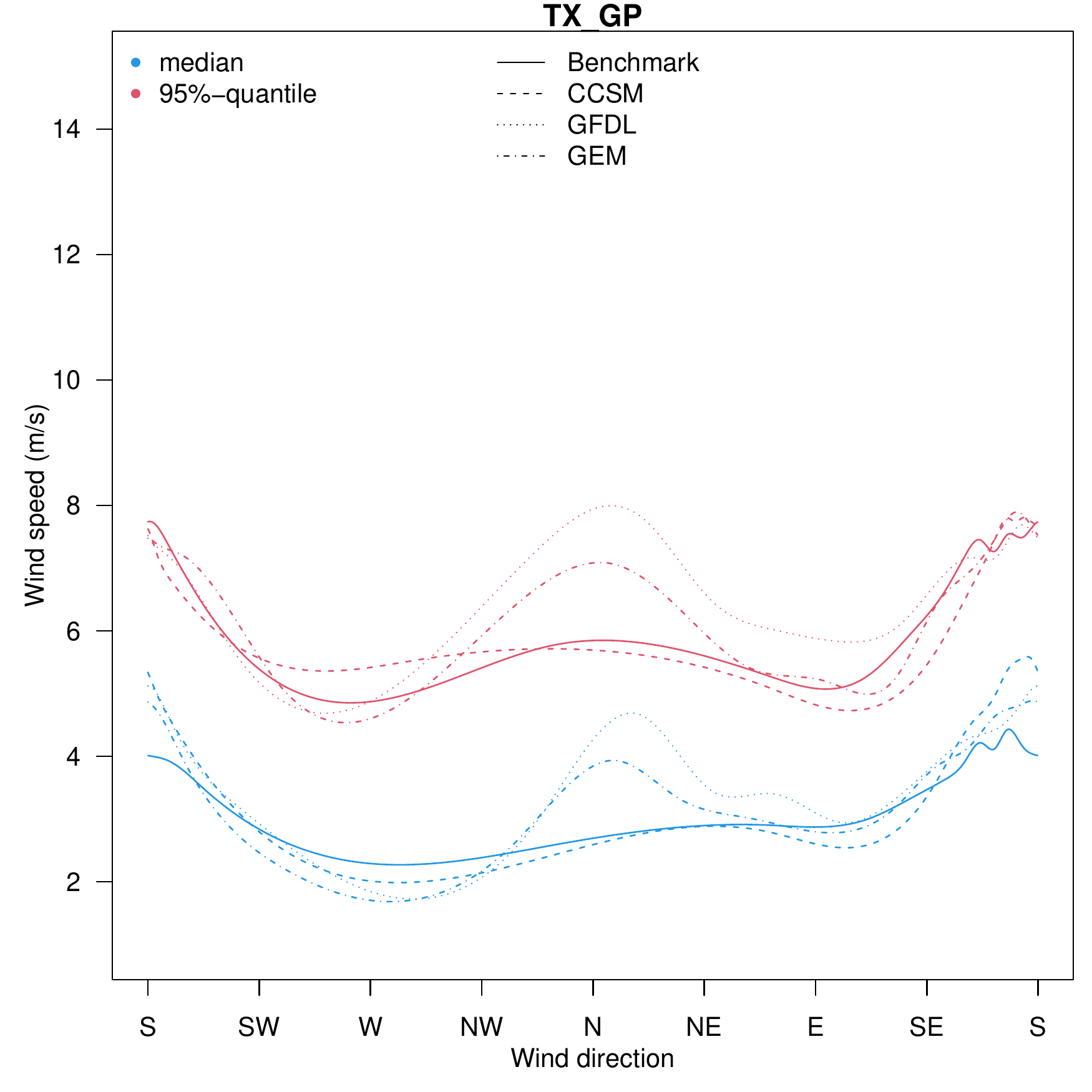}
 \caption{50th (blue) and 95th (red) quantiles of wind speed as a function of wind direction from the quantile regression model for historical period in summer for three locations, \textbf{from left to right}: offshore California, Colorado and Texas-SGP. The different RCMs are shown with different types of dotted and dashed line, benchmark data are in solid lines (NARR for offshore and NLDAS for inland).}
 \label{fig:qr1.sum} 
 \end{figure}

We use the quantile regression with periodic B-splines to estimate the median and 95th-quantile of the directional wind speed. 
Figure \ref{fig:qr1.win} compares the quantile regression results using WRF and NLDAS over CO and Texas-SGP inland locations, and WRF and NARR at the CA location in winter. (Results for the remaining locations can be found in supplementary materials.) 
For the offshore location on CA coast (left panel), both median and the 95th-quantile in the benchmark data NARR show the strongest wind speed concentrated in the northwest directions, favorable for stable wind resources. All three WRF models capture high wind speeds in west and northwest direction but also generates high wind speeds in northeast and east directions that are not present in the NARR data. In addition, three WRF simulations tend to overestimate wind speed compared with NARR and NLDAS for all the directions, and all other offshore locations are also overestimated by WRF. According to the validation of NARR using buoy in Figure \ref{fig:diurnal1}, the apparent overestimation could mostly be because of the systematic low bias of NARR. For example, over the CA coast, the wind speed is underestimated in NARR by about 4m/s. Therefore, the overestimation seen in Figure \ref{fig:qr1.win} and Figure \ref{fig:qr1.sum} over CA coast is not as concerning as it appears. 

Over the Colorado location, WRF simulations capture the highest wind speed in south and southwest directions and the lowest wind speed in northeast and east. 
Over the Texas location, WRF simulations capture the highest wind speed in northwest and north, and also the magnitude of wind speed, indicating that reanalysis benchmark data and WRF simulations show a strong consistency. 
Note that winds are relatively easier to simulate in flat regions than over complex terrains such as mountains and coastal lines. 
WRF-simulated dominant wind directions over Texas and  Colorado locations are overall similar to that from NLDAS. 

Figure \ref{fig:qr1.sum} shows the quantile regression results for WRF and benchmark data over the same location as shown in Figure \ref{fig:qr1.win}, but in summer. While the two inland locations have much weaker wind speeds in summer than in winter, the offshore location over CA shows similar high wind speed (95th-quantile), however its median wind speed is stronger in summer than in winter. 
WRF outputs capture the dominant wind direction over all three locations. 
Interestingly, the highest wind speeds over all three locations are located in the same direction as seen in winter. For example, over the offshore location in CA coast, the highest wind speed concentrated over the northwest locations. Over the inland locations, the highest wind speed is from south and southwest directions over the Colorado location, and north and south over the location in Texas. This is perhaps encouraging information to wind energy resource development, as stable wind directions can ensure stable wind energy production. 
One potential reason for the inconsistency could be the resolution difference, as NARR has 32km-resolution while WRF has 12km-resolution. A mismatch in co-location at the grid-point level is typically more visible for offshore locations. Another reason for the direction bias could be that mountain areas are more sensitive to the wind direction.

\section{Future projections of wind conditions}\label{sec:projection}

In this section, we focus on projections of wind conditions in late-century (2085 - 2094) under the scenario of RCP 8.5 compared with wind conditions in historical period (1995-2004). We also conducted the same analysis using mid-century projections, and the climate change signal is smaller but the conclusions are qualitatively similar (not shown). The changes in directional wind speed distributions are estimated using   
quantile regression and Weibull distributional regression where bootstrap (Sec.~\ref{sec:boot}) are used to quantify the estimation uncertainty. We also compute their projected changes for the standard deviation and 95th-quantile of wind speeds, and compare with them with the corresponding statistics of internal variability defined in Sec. \ref{sec:iv}. 
Finally, in addition to the near-surface wind, we also discuss the uncertainty due to internal variability versus projected changes in wind conditions at higher heights relevant to wind industry (up to 200 meters) in WRF outputs to shed some light on potential future changes in wind energy resources.

\subsection{Changes in wind speed and direction distributions}

\begin{figure}
 \centering 
 \includegraphics[scale= 0.35]{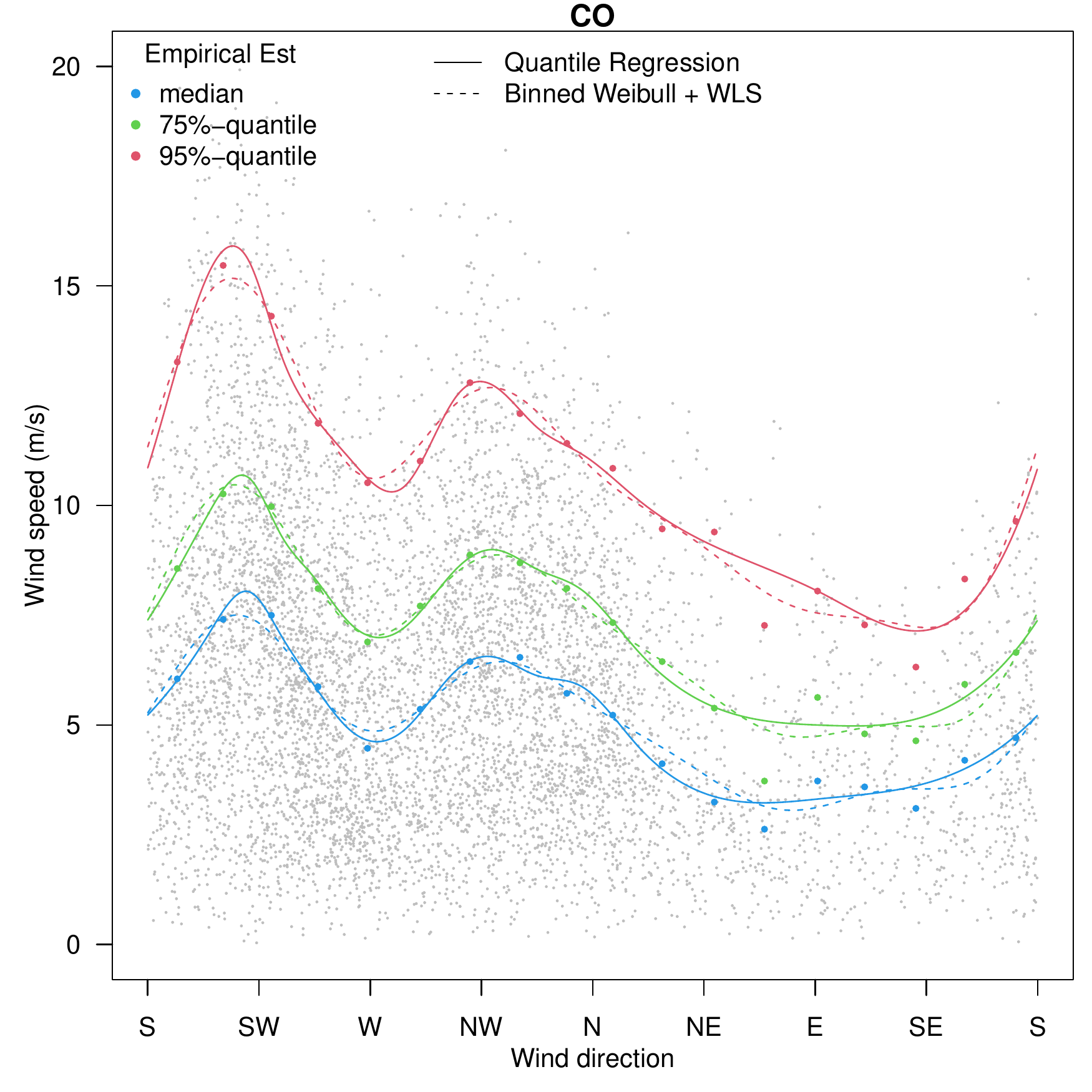} 
  \includegraphics[scale= 0.35]{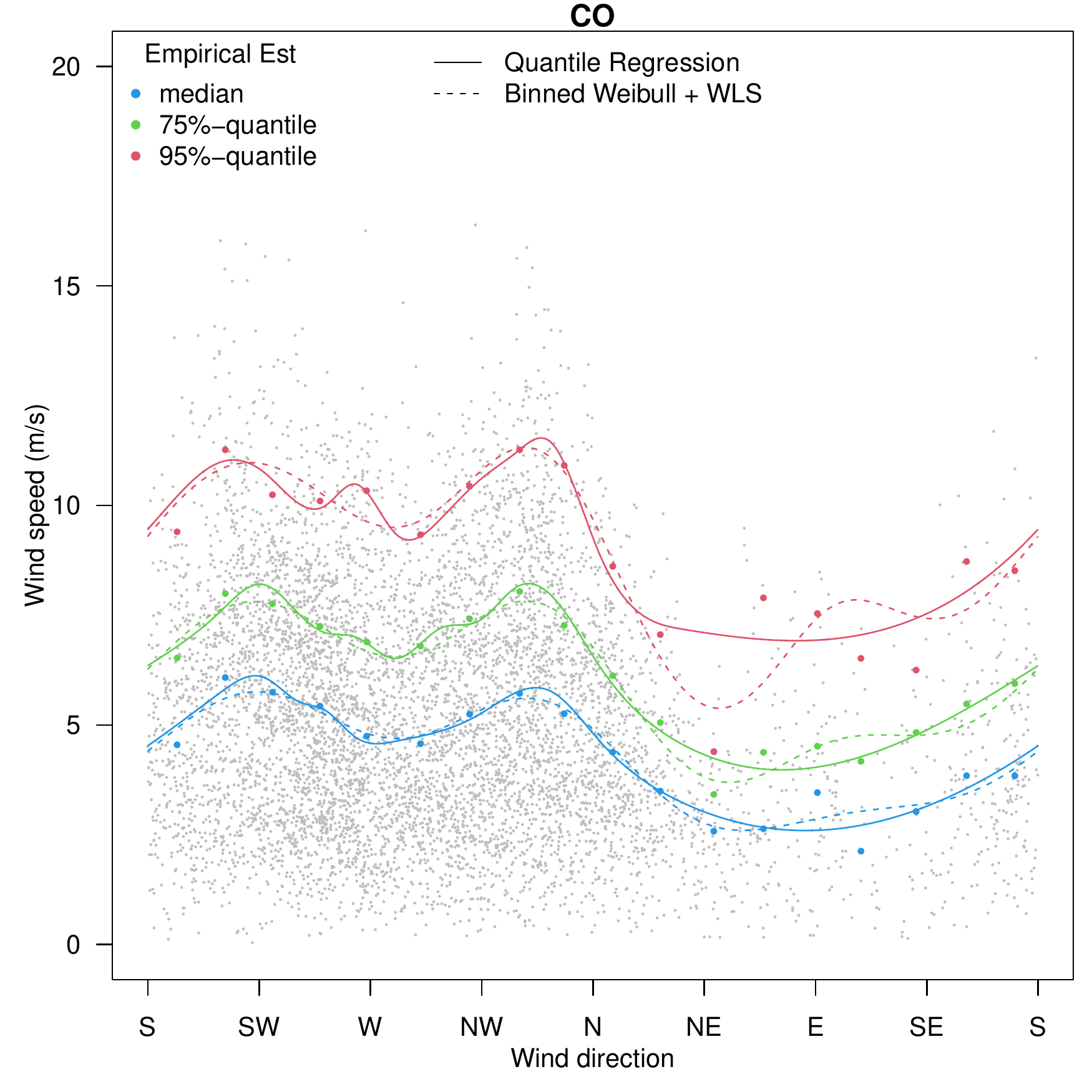} \\
  \includegraphics[scale= 0.35]{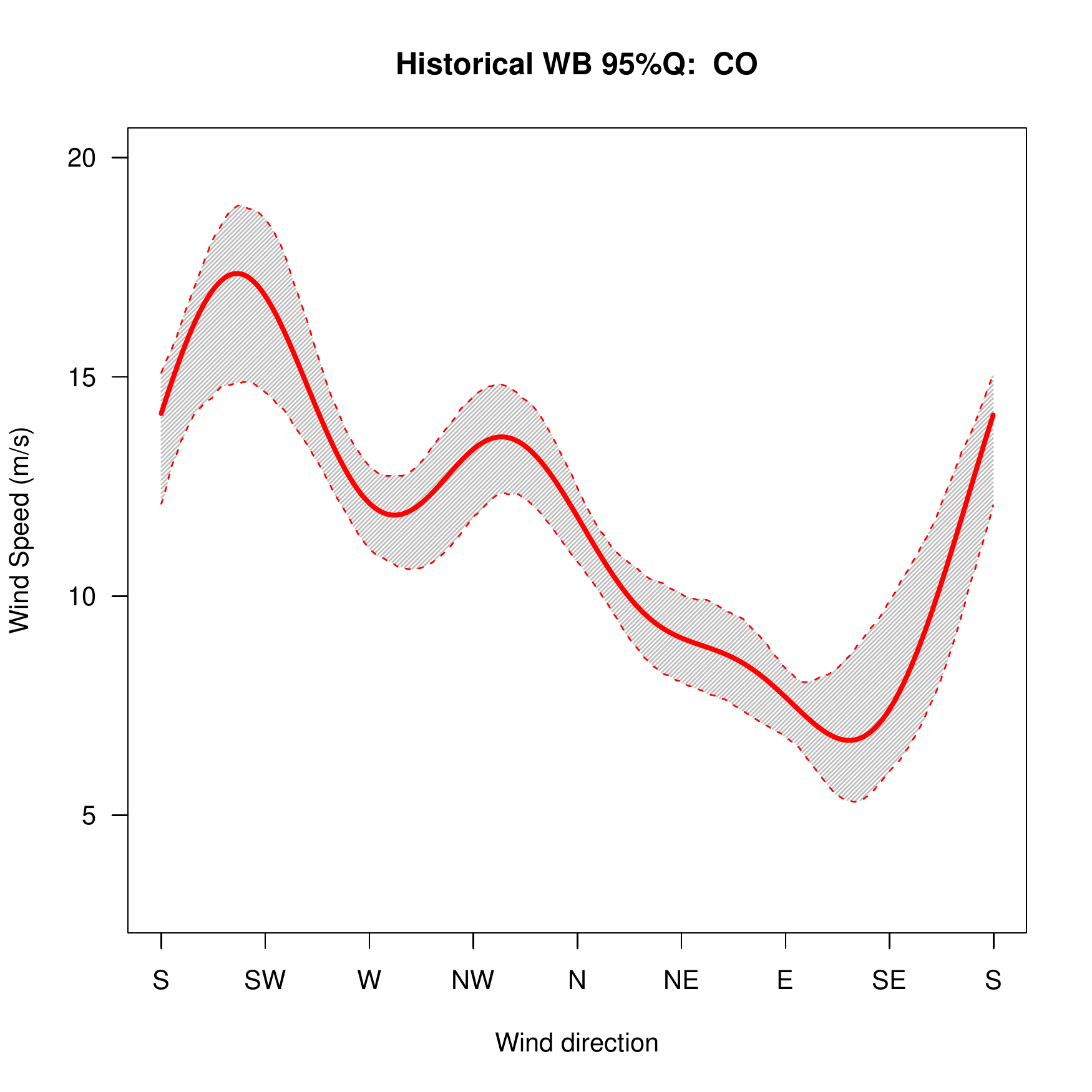}
  \includegraphics[scale= 0.35]{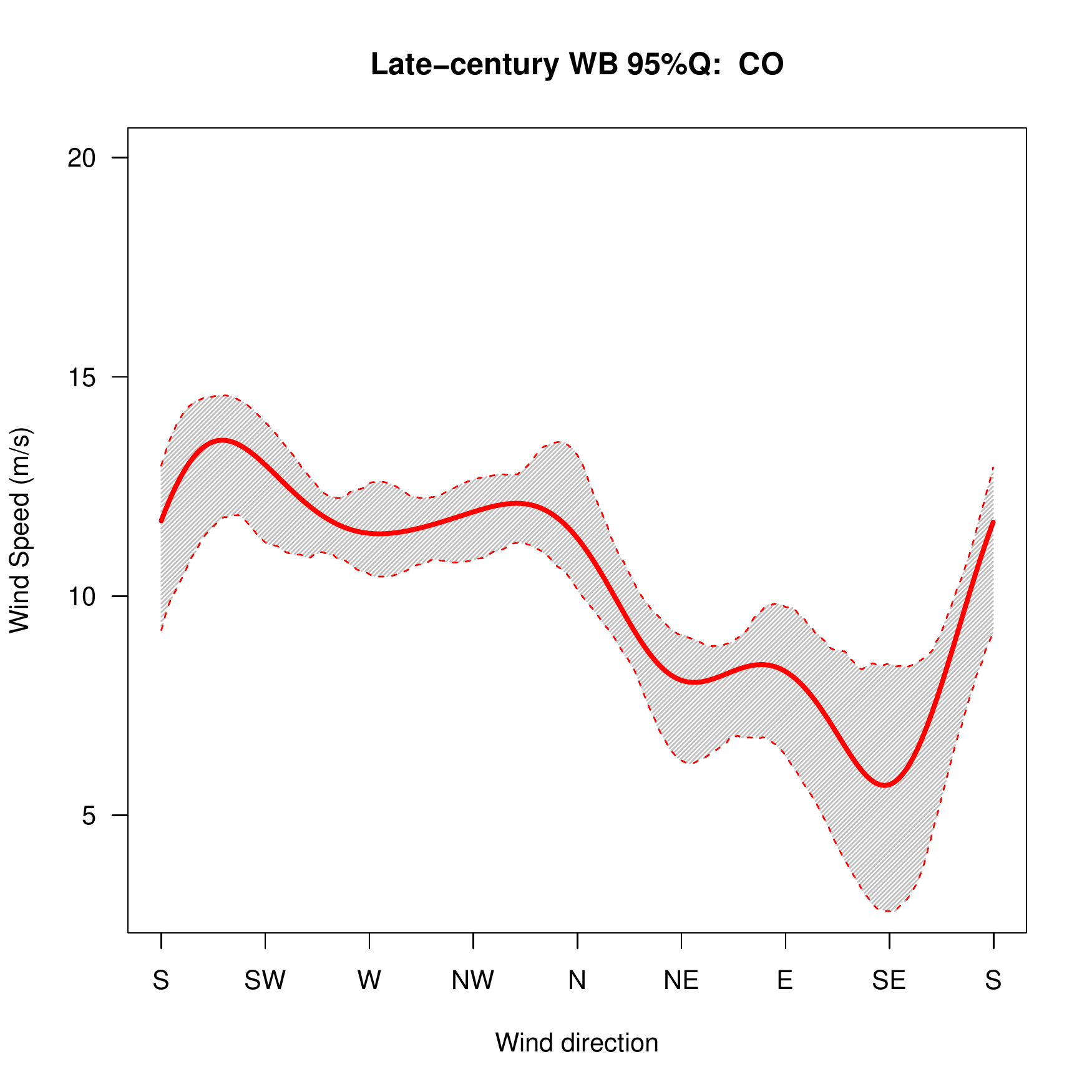}
 \caption{Estimated directional quantiles (\textbf{upper panels}, blue for 50th, green for 75th, and red for 95th-quantile) and their bootstrapping estimation uncertainties (\textbf{lower panels}, only results for the 95th-quantile are shown) at the CO mountain grid-cell for winter season under the historical (\textbf{left}) and late-century (\textbf{right}).}\label{fig:qr13}
 \end{figure}

The change in wind direction estimated by von Mises distribution between historical period and late-century period is fairly small in all the scenarios as shown in Sect. S1 of the supplemental material, thus, in the following, we focus on the change in the directional wind speed.  
Figure \ref{fig:qr13} shows the estimated directional wind speed quantiles (50th, 75th and 95th) for WRF-HadGEM outputs at the CO mountain location in winter using both quantile regression and Weibull distributional regression (upper panels), and the associated 95\% pointwise bootstrapped confidence interval (for 95th-quantile) using Weibull distributional regression. The corresponding results for the remaining 9 locations can be found in Sect. 2 of the supplemental material. 
Compared with the directional wind speed quantile estimates from historical period, the projected winds at CO mountain location show evident weaker directional wind speeds, especially for the dominant wind direction southwesterlies, but also other directions such as westerlies, and northwesterlies from both quantile regression and Weibull distributional regression models at all three quantiles. 

We also find the dominant directional wind speeds decrease at the Idaho inland location, North Dakota and FL, as shown in Sect. 2 of the Supplementary Materials. In addition to wind speed change at the dominant wind directions, there are changes in future directional wind speed pattern. For example, in Oregon coastal location during the historical period, the northwesterly winds dominant both median and high wind speeds; in projections, there are stronger winds from the southeast direction as well. Over Lake Erie, in historical periods, dominant wind speeds are southwesterly and northeasterly; in future, the northeasterly wind is decreasing while the northerly wind is increasing. 
 In summary, the change in median winds from historic to late-century is noticeable in some locations such as CO mountain and Idaho inland, but minor in most locations. However, the 95th-quantile directional wind speeds tend to show more changes.

\subsection{Projected climate change versus internal variability}
From previous discussions we noticed that several locations may show decrease in wind speed at their dominant wind directions, while other locations may experience changes in dominant wind directions themselves, yet other locations do not exhibit any significant wind condition changes in late-century. In this section, we investigate the robustness of such projected changes by considering the uncertainty due to internal variability \citep{wang2018} that is caused by a perturbation in initial conditions. 
We calculate the commonly used internal variability as well as the newly defined internal variability for standard deviation (left panels) and 95th-quantile (right panels) following Sect. \ref{sec:iv}. 
These latter enable the assessment of the significance of projected changes over the intrinsic variability of the models for other statistics (standard deviation and 95th-quantile) than the commonly used mean statistics. 

\begin{figure}
 \centering
  \includegraphics[scale= 0.55]{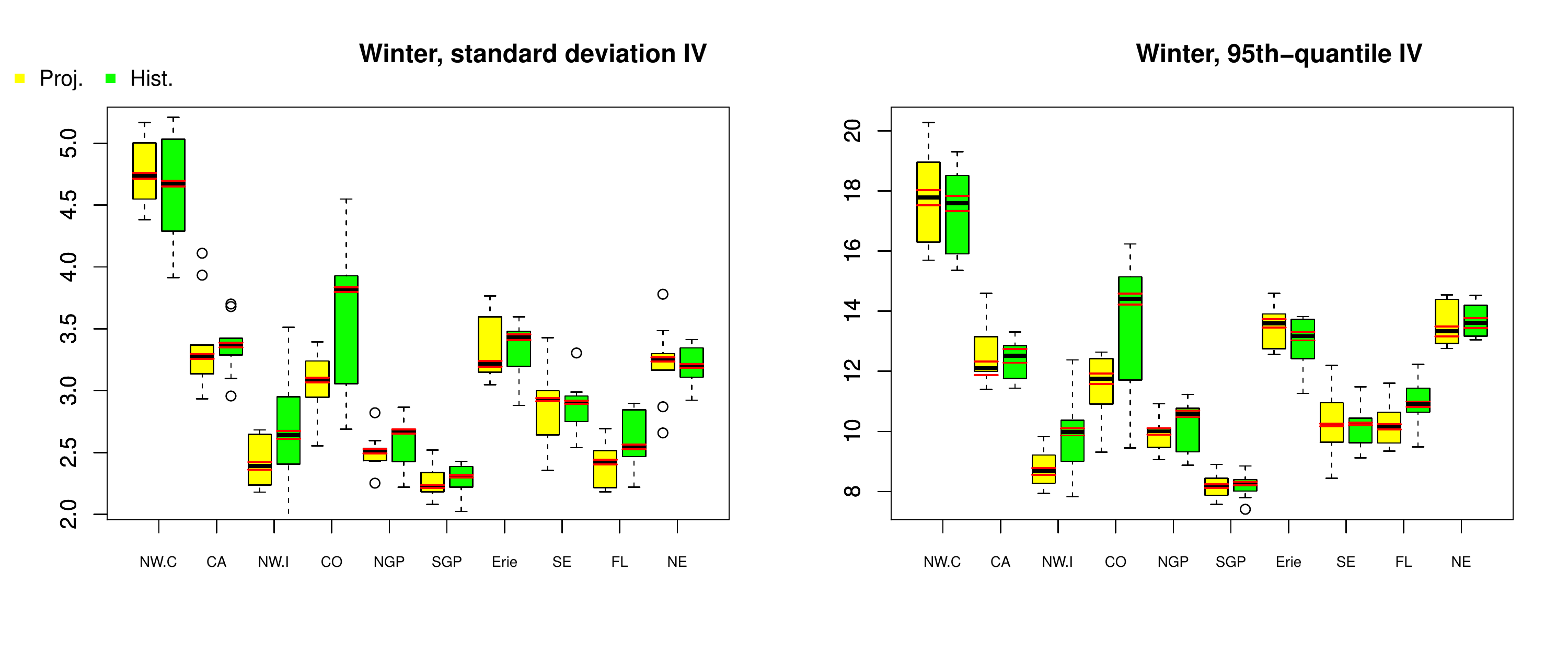}
   \includegraphics[scale= 0.55]{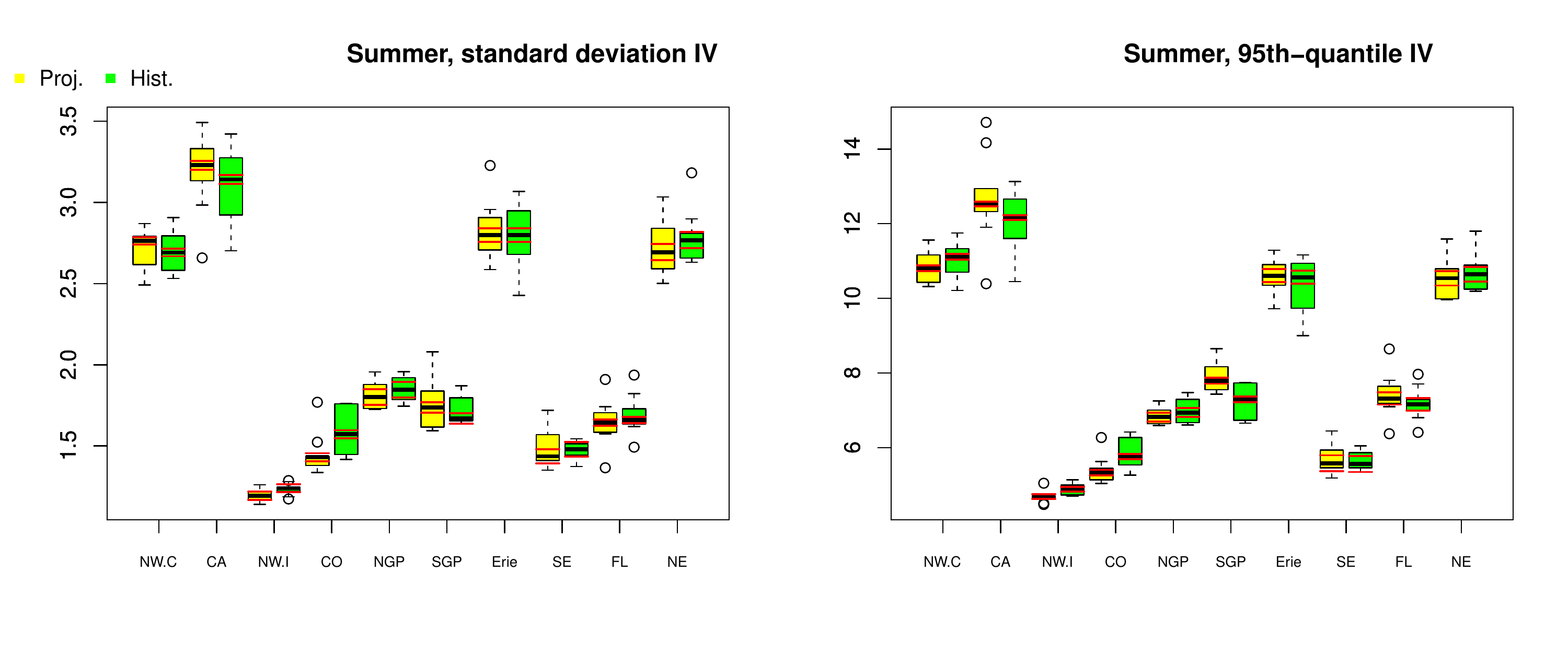}
 \caption{Boxplot of standard deviation (\textbf{left}) and 95th-quantile (\textbf{right}) yearly PCC variability with  corresponding IV in 10 locations from WRF-CCSM. The boxes and whiskers are 1-year based statistics for 10 years in historical and future periods, indicating the yearly PCC variability of each 10-year period. The red lines represent the median statistic +/- the IV for each location.  If the difference between two means from historic and late-century are larger than the IV, then we consider that the future changes in these statistics are robust. }\label{fig:pcc/iv}
 \end{figure}

In the following, in order to generate observe variability of PCCs and gather more information on PCC than a single PCC statistic (difference between historical and projected statistic), we compute the PCCs are differences of yearly statistics. 
Figure \ref{fig:pcc/iv} presents the yearly PCC variability (boxplot) 
and the internal variability (red lines) over the 10 studied locations across the US for standard deviation and 95th-quantile in winter and summer. First, we calculate the standard deviation and the 95th-quantile of each year in the historical (1995-2004) and future (2085-2094) periods, and consider all the differences between historical and projected statistics. The boxplot (box, whiskers, and black line for the median) shows the distribution of these differences (yearly historical and yearly projected) for seasonal standard deviation and seasonal 95th-quantile. 

In each case, the internal variability is added and subtracted to the median values of standard deviations and 95th-quantiles. When the difference between the median of historical and future is two times larger than the internal variability, then we consider the climate change of standard deviation and 95th-percentile of wind speed is significant with respect to the internal variability.  

We observe that, in winter, the standard deviation is reduced in late-century over seven locations, especially over the Northwest inland location (location 2), CO mountain (location 4), and Lake Erie (OH location 10), where the standard deviations are reduced robustly, indicating the variability of winter wind speed in projections over these locations are smaller or simply the wind speed in future winter is weaker. 
The projected high wind speed (95th-percentile) over these locations also show a robust decrease (top right panel), suggesting lesser wind energy resources in the future. In summer, both the yearly PCC variability and the projected changes in standard deviation and 95th-quantiles are relatively small compared to those statistics in winter. There are slight increases in standard deviation and wind speed 95th-quantile over CA offshore (location 1) and Texas Great Plain (location 5), however these increases are not significant compared with the internal variability.  

To explore future changes in wind speeds at higher heights (e.g, wind turbine's hub-heights), we have conducted the same analysis for three higher heights above the ground level, at 28.48, 97.88, 192.39 meters from the WRF runs. We found similar conclusions to the near surface wind speed (results are not shown). That is, while the mean wind speed change is not robust with respect to internal variability, the changes in standard deviation and 95th-quantile are robust over certain locations in winter. For example, the wind speed is projected to decrease over Northwest inland and Colorado mountains. 

\section{Summary and discussion}\label{sec:conclusion}

In this work, we study wind conditions and their potential projected changes via a statistical conditional framework, where probability distributions of wind directions are modeled via von Mises mixture distributions, and directional wind speed distributions are characterized by two distributional regression models, namely quantile regression and Weibull regression models. The proposed framework allows us to better characterize wind direction distributions, directional wind speed distributions, and hence provides the full description of joint wind speed and direction distributions. In addition we investigate the strength of the projected wind speed distributional changes relatively to the interval variabilities of the climate models.  The extension beyond the mean change to the changes and internal variability of standard deviation and 95th-quantile provides a more complete picture on assessing the significance of their changes relatively to their internal variability.

We also perform a comprehensive climate model evaluation where RCM outputs are evaluated against reanalysis data and in-situ measurements. Results from the WRF configurations are fairly consistent in both wind direction and speed. Wind direction and speed in summer are generally less dispersed than the winter scenarios. We observe that the benchmark NARR and NLDAS data, corroborated by buoy and ground-station data, are mostly consistent with the WRF outputs in most locations.
Our evaluation study also highlights the challenges of finding appropriate benchmark data for these high-resolutions RCMs in some situations (e.g., offshore, mountain areas). Few data products are available at high resolution and in-situ measurements are irregularly available in space with varying time resolutions. 

This study concludes that the changes between historical and future wind directions are usually small in the locations we examined, but wind speeds are generally weakened in most locations we considered in the projected period, with some locations and directions get intensified in the future. 
The projected climate changes in 95th-quantile and standard deviation are significant over the internal variability at some locations, and decrease in locations such as northwest inland, Colorado, and Lake Erie, especially in the winter season.

The current implementation of our statistical framework has some limitations. 
First, both distributional regression methods require some ``tuning'': the Weibull regression requires choosing the binning and determining the complexity regression functional form (via the number of the harmonic terms), whereas the quantile regression requires to select knots and degree of freedom  of periodic B-splines. Further statistical studies of these modeling choices are underway and should be available in near future.
Second, all the statistical analyses  are performed pointwise for simplicity, spatiotemporal dependent structures are  not considered here. In order to assess regional wind fields and their future predictability it is critical to explore their spatiotemporal structures in future work.

\appendix

\section{Quantile regression method} \label{sec: quantile regression}
In this work we model the $\tau$-quantile of wind speed (\texttt{WS}) conditioning on wind direction  (\texttt{WD}) using a quantile regression with periodic B-spline. The estimator of $ Q_{\texttt{WS}|\texttt{WD}}(\tau| \texttt{wd}) $ takes the following form
\begin{equation}\label{eq: QR}
    \hat{Q}_{\texttt{WS}|\texttt{WD}}(\tau| \texttt{wd}) = Z(\texttt{wd})^\top \hat{\beta}(\tau). 
\end{equation}
The vector $Z(\texttt{wd})$ is a periodic B-spline with degree of freedom $s$ evaluated at wind direction value $\texttt{wd}$. The periodic B-spline, instead of more commonly used B-spline, is used here to preserve the periodic directional property of the directional wind speed distribution. 
We select the degree of freedom  ($s=8$) by finding the elbow point of the mean absolute error (MAE) of the regression residuals versus degree of freedom ratio. 
The coefficient vector $\hat{\beta}(\tau)$ is the QR estimator of $\beta(\tau)$, i.e.:
\begin{equation}\label{eq: QR2}
     \hat{\beta}(\tau) = \arg\min_\beta \sum_{i=1}^n\rho_{\tau}\left(\texttt{ws}_i - Z(\texttt{wd}_i)^\top \beta(\tau)\right). 
\end{equation}
%
where $\rho_{\tau}$ is a loss function for $\tau$th quantile. It usually can be expressed as 
\begin{equation}\label{eq: QR3}
     \rho_{\tau}(y) = y(\tau - \mathbb{I}(y<0)). 
\end{equation}
We seek to minimize quantile loss by differentiating Equation \ref{eq: QR4} with respect to $\hat{y}$
\begin{equation}\label{eq: QR4}
     E_{\rho_{\tau}}(Y-\hat{y}) = (\tau -1)\int_{-\infty}^{\hat{y}}(y-\hat{y})dF(y) + \tau\int^{\infty}_{\hat{y}}(y-\hat{y})dF(y). 
\end{equation}

\section{Acknowledgement}
We acknowledge the support from AT\&T Services Inc. under a Strategic Partnership Project agreement A18131 to Argonne National Laboratory through U.S. Department of Energy contract DE-AC02-06CH11357. We acknowledge the National Energy Research Scientific Computing Center (NERSC), Argonne's Laboratory Computing Resource Center (LCRC), and the Argonne Leadership Computing Facility (ALCF) for providing the computational resources used to

\bibliographystyle{apalike}
 \bibliography{references.bib}
\end{document}